\documentclass[prd,preprint,tightenlines,floats,floatfix,showpacs,showkeys,preprintnumbers,nofootinbib,eqsecnum]{revtex4}
 \usepackage[dvips]{graphicx}
  \usepackage{amsmath}
   \usepackage{amssymb}
    \usepackage{pifont}

\newcommand{\Ds}{\displaystyle}                                   
\newcommand{\va}[1]{\langle{#1}\rangle}                           
\newcommand{\gev}[1]{\relax\ifmmode{\text{GeV}^{#1}}\else{GeV$^{#1}${ }}\fi}
\newcommand{\Gev}{\relax\ifmmode{\text{GeV}}\else{GeV{ }}\fi}     
\newcommand{\Mev}{\relax\ifmmode{\text{MeV}}\else{MeV{ }}\fi}     
\def\MS{$\overline{\text{MS}\vphantom{^1}}${ }}
\def\muF{\relax\ifmmode\mu_\text{F}^2\else{$\mu_\text{F}^2${ }}\fi}
\def\as{\relax\ifmmode \alpha_s\else{$ \alpha_s${ }}\fi}          
\newcommand\convo[1]{\mathop{\otimes}\limits_{#1}}                

\begin{document}
\thispagestyle{empty}
\date{\today}

\preprint{JINR-E2-2005-21, RUB-TPII-10/04}
\title{Renormalization-group improved evolution of the meson
       distribution amplitude at the two-loop level}

\author{Alexander~P.~Bakulev}
 \email{bakulev@thsun1.jinr.ru}
\affiliation{%
  Bogoliubov Laboratory of Theoretical Physics,  \\
  Joint Institute for Nuclear Research,          \\
  141980, Moscow Region, Dubna, Russia}%

\author{N.~G.~Stefanis}
 \email{stefanis@tp2.ruhr-uni-bochum.de}
\affiliation{%
  Institut f\"ur Theoretische Physik II          \\
  Ruhr-Universit\"at Bochum                      \\
  D-44780 Bochum, Germany}%
\vspace {10mm}

\begin{abstract}
We discuss the two-loop evolution of the flavor-nonsinglet meson
distribution amplitude in perturbative QCD.
After reviewing previous two-loop computations, we outline the
incompatibility of these solutions with the group property of the
renormalization-group transformations.
To cure this deficiency, we compute a correction factor for the
non-diagonal part of the meson evolution equation and prove that
with this modification the two-loop solution conforms with the
group properties of the renormalization-group transformations.
The special case of a fixed strong coupling (no $Q^2$ dependence)
is also discussed and comparison is given to previously obtained
results.
\end{abstract}

\maketitle

\newpage
\section{Introduction}
\label{sec:intro}
In order to use perturbative QCD (pQCD) for the description of
exclusive processes, one usually appeals to factorization theorems,
which ensure that the pQCD calculable information, described by the
so-called hard scattering amplitude, $T_{\rm H}$, can be factored out
at a factorization scale $\muF$, whereas all pQCD non-calculable (i.e.,
non-perturbative) contributions are encoded in terms of universal
hadron distribution amplitudes (DA)s, $\varphi_{h}$.
Schematically, this can be illustrated for the case of the
$\gamma^*\gamma^*\to\pi$ transition form factor in the following way:
\begin{eqnarray}
 \label{eq:FF-factorization}
  F_{\gamma^*\gamma^*\to\pi}(Q^2,q^2)
   = f_{\pi}
     T_{\rm H}(x;Q^2,q^2,\muF)
      \convo{x}
       \varphi_{\pi}(x;\muF)\,,
\end{eqnarray}
where $Q^2$ and $q^2$ are the photon virtualities and $\otimes$
denotes the usual convolution symbol
($A(x)\convo{x}B(x) \equiv \int_0^1 dx A(x) B(x)$)
over the longitudinal momentum fraction variable $x$.
The dependence of the DA $\varphi_{\pi}(x;\muF)$ on the factorization
scale $\muF$ is governed by the Efremov--Radyushkin--Brodsky--Lepage
(ERBL) evolution equation \cite{ER80,LB79}
\begin{eqnarray}
 \frac{d\,\varphi_{\pi}(x;\muF)}
      {d\,\ln\muF}
  &=& V(x,u;\alpha_s(\muF))\convo{u}\varphi_{\pi}(u;\muF)
 \label{eq:ERBL}
\end{eqnarray}
with the evolution kernel
\begin{eqnarray}
 V(x,u;\alpha_s(\mu^2)))
  &=& \left(\frac{\alpha_s(\mu^2))}{4\pi}\right)V_0(x,y)
  + \left(\frac{\alpha_s(\mu^2))}{4\pi}\right)^2V_2(x,y)
  + \ldots\,,
 \label{eq:kernel}
\end{eqnarray}
adopting the notation of \cite{MR86ev}.
The one-loop evolution kernel $V_0$ was introduced in Ref.\
\cite{LB79}; an analogous expression for $V_2$ at the two-loop level
was derived in Refs.\ \cite{DR84,Sar84,Kat84,MR85}.
If factorization applies, it is, in principle, safe and legitimate
to use different values of $\muF$, dispensing with the
factorization-scale dependence by means of the renormalization group.
Usually, one sets $\muF=\bar{Q}^2\equiv Q^2+q^2$ in order to eliminate
in $T_{\rm H}$ large logarithms of the type $\ln(\bar{Q}^2/\muF)$ and
for that reason one then needs to evolve the meson DA
$\varphi_{\pi}(x;\mu^2)$ in accordance with Eq.\ (\ref{eq:ERBL})
from a low normalization point $\mu_0^2\lesssim1$~GeV$^2$,
at which the pion DA has been (non-perturbatively) determined,
to a higher scale $\mu^2=\bar{Q}^2$ at which comparison with
experimental data may be attempted.

The structure of the two-loop evolution kernel was analyzed by
Mikhailov and Radyushkin (MR) in~\cite{MR86ev} and its diagonality
violating terms (in the Gegenbauer basis) were identified, though
their origin remained partly unclear.
Moreover, the two-loop evolution with $Q^2$ of any pion DA at the
reference (or initial) momentum $\mu_0^2$ was carried out numerically,
but including only the first few expansion coefficients at order
$\alpha_s^2$.
On that basis, these authors concluded that the next-to-leading-order
(NLO) correction to the pion DA with $\mu_0^2\simeq1$~GeV$^2$ and at
$x=0.5$ remains less than a few percent even for momenta far beyond
$10^2$~GeV$^2$.
Later, the two-loop evolution of the pion DA was studied by M\"{u}ller
in \cite{Mul94} using conformal constraints.
In this work, a complete formal solution in NLO was obtained with the
inclusion of all two-loop mixing coefficients.
This solution was further discussed in~\cite{Mul95} for a running and
also a fixed coupling constant.
It was found that the NLO correction can be rather large -- especially
for endpoint-concentrated DAs, like the Chernyak--Zhitnitsky (CZ)
\cite{CZ84} one, supplying logarithmic enhancement exactly in this
region.

However, as we will show below, all these solutions violate the group
character of the renormalization-group (RG) evolution transformations
to the order $\alpha_s^2$.
Whether it is possible to obtain an approximate solution of the
evolution equation that, nevertheless, respects its group properties,
remains an open question.
It is exactly this issue to which the present work is devoted.
Such an improvement is not only theoretically important, it is also of
practical concern since the modification of the evolution behavior,
entailed by the restoration of the RG properties, will have influence
on measurable hadronic observables, like meson form factors.

In the next section, we shall briefly review the main properties of the
two-loop approximation of the QCD coupling and describe the standard
formalism for the DA evolution at the two-loop level.
Section \ref{sec:RGPro-2} presents the analysis of the RG
transformation when the second and higher Gegenbauer harmonics are
taken into account.
A semi-explicit solution of the two-loop evolution equation that
preserves the RG structure is constructed in Sec.\ \ref{sec:RG-Ex}.
In Sec.\ \ref{sec:Numerical} we discuss the numerical importance of
the obtained solution.
The special case of a fixed coupling constant is treated in Sec.\
\ref{sect:RG_Fixed}, while some important technical details are
collected in two appendixes.

\section{QCD coupling and evolution of the pion DA at NLO}
 \label{sec:2-loop_alpha}
  \subsection{Standard formalism with standard notations}
To illustrate the ``RG philosophy'' and clarify the objective of this
work, let us recall the Ovsyannikov--Callan--Symanzik equation for the
(running) coupling, $\alpha_s\left(Q^2\right)$, in QCD:
\begin{eqnarray}
 \label{eq:RG-coupling}
  \frac{d \alpha_s\left(Q^2\right)}{d \ln(Q^2)}
   = \beta\left(\alpha_s(Q^2)\right)\,.
\end{eqnarray}
Strictly speaking, $\alpha_s(Q^2)$ and also $\beta(\alpha_s(Q^2))$
depend on the number of active flavors, $N_f$.
For the considerations to follow, this will not be important and hence
we omit the $N_f$-dependence any further.
The $\beta$-function in the NLO approximation is given by
\begin{eqnarray}
 \label{eq:RG-beta}
  \beta\left(\alpha_s\right)
  = -\frac{\alpha_s^2}{4\pi}
      \left(b_0 + b_1 \frac{\alpha_s}{4\pi}
      \right)\, ,
\end{eqnarray}
where the standard $\beta$-function coefficients are provided in
Appendix~\ref{sec:App-A}.
The two-loop equation for $\alpha_s(Q^2)$
(with $\Lambda_\text{QCD}\equiv \Lambda$)
reads
\begin{eqnarray}
 \label{eq:als_2-loop}
  \frac{4\pi}{b_0\alpha_s(Q^2)} -
   c_1
    \ln\left[\frac{4\pi}{b_0\alpha_s(Q^2)} + c_1
       \right]
  = \ln\left(\frac{Q^2}{\Lambda^2}\right)
 \quad\text{with~}
 c_1\equiv b_1/b_0^2 \,.
\end{eqnarray}
The solution of this equation can be written in terms of the Lambert
$W_{-1}$-function with the argument
$ \zeta(Q^2)
\equiv
  -\frac{1}{e c_{1}} \left(\frac{\Lambda^{2}}{Q^{2}}\right)^{1/c_{1}}$
to obtain
\begin{eqnarray}
 \label{eq:Lambert}
  \as^\text{2-loop}\left(Q^2\right)
  &=& -\frac{4\pi}
     {b_0\left(N_f\right)c_1
      \left[1 + W_{-1}(\zeta(Q^2))\right]}\,,
\end{eqnarray}
as it was shown in \cite{Mag98,GGK98}.

Now let us turn to the ERBL\ evolution equation for the meson DA.
First, we recall the main properties of the one-loop approximation.
(See Ref.\ \cite{Ste99} for a pedagogical exposition and further
references.)
The one-loop evolution kernel has a very simple structure, viz.,
\begin{eqnarray}
 \label{eq:V_ERBL_1-loop}
  V_\text{1-loop}(x,y;\alpha_s)
  &=& \frac{\alpha_s}{4\pi} V_0(x,y)
\end{eqnarray}
with factorizing $\alpha_s$- and $x$-dependences.
The formal solution of the ERBL\ equation in the one-loop approximation
is
\begin{eqnarray}
 \label{eq:EvPDA}
  \varphi_\pi(x;\mu_0^2)
   \stackrel{\text{ERBL}}{\longrightarrow}
    \varphi_\pi^\text{1-loop}(x;Q^2)
  &=&\exp
      \left[\int_{\alpha_s(\mu_0^2)}^{\alpha_s(Q^2)}
       \frac{\alpha_s V_0}{4\pi\beta_1(\alpha)}d\alpha
        \convo{}
      \right]
      \varphi_\pi(x;\mu_0^2)\,,
\end{eqnarray}
where $\beta_1(\alpha)=-b_0 \alpha_{s}^{2}/(4\pi)$ is the one-loop
$\beta$-function and the exponent above has to be evaluated according
to
$\Big[V_0\convo{}\Big]^{n}\varphi_\pi(x;\mu_0^2)
= V_0(x,u_1)\ \convo{u_1}\
   \ldots\convo{u_{n-1}}
    V_0(u_{n-1},u_n)\convo{u_n}
     \varphi_\pi(u_n;\mu_0^2)$.
It is useful to expand the meson DA\footnote{%
Here $\Ds{\sum_{n>0}}'$ denotes the sum over even indices $n>0$ only in
order to account for the symmetry relation
$\varphi_\pi(x;Q^2)=\varphi_\pi(1-x)$ due to charge-conjugation
invariance and isospin symmetry.}
\begin{eqnarray}
 \label{eq:Gegenbauer}
  \varphi_\pi(x;Q^2)
  &=& \Omega(x){\sum_{n\geq0}}'a_{n}(Q^2)\cdot \psi_{n}(x)
\end{eqnarray}
in terms of the eigenfunctions $\Omega(x)\psi_n(x)$ of the one-loop
ERBL\ kernel (\ref{eq:V_ERBL_1-loop}), i.e., in terms of the Gegenbauer
polynomials $C^{3/2}_n(\xi)$,
\begin{eqnarray}
 \Omega(x)&\equiv& 6\, x (1-x)\,,\qquad
 \psi_n(x)\ \equiv\ C^{3/2}_n(2x-1)\,,
\end{eqnarray}
corresponding  to the eigenvalues
\begin{eqnarray}
  \gamma_n^\text{1-loop}(\alpha_s)
  = \frac{-1}{2}
      \left(\frac{\alpha_s}{4\pi}\right)\gamma_0(n)
\end{eqnarray}
with $\gamma_0(n)$ given in Appendix~\ref{sec:App-A}.
In this representation all the dependence on $Q^2$ is contained in the
coefficients $a_n(Q^2)$:
\begin{eqnarray}
 \label{eq:1-loop_a_n}
  a_n^\text{1-loop}(Q^2)
   &=& a_n(\mu_0^2)
       \left[\frac{\alpha_s(Q^2)}{\alpha_s(\mu_0^2)}
       \right]^{\gamma(n)}\,
  \quad \text{with}\quad
  \gamma(n)\ \equiv\ \frac{\gamma_0(n)}{2b_0}\,.
\end{eqnarray}
This simple scheme is due to the special factorized structure
of the one-loop evolution kernel (\ref{eq:V_ERBL_1-loop}).
In the two-loop approximation this is no more the case.
Indeed, the evolution kernel
\begin{eqnarray}
 V_\text{2-loop}(x,y;\alpha_s(\mu^2))
  &=& \left(\frac{\alpha_s(\mu^2)}{4\pi}\right)V_0(x,y)
  + \left(\frac{\alpha_s(\mu^2)}{4\pi}\right)^2V_2(x,y)
 \label{eq:V_ERBL_2-loop}
\end{eqnarray}
has, at each scale $\mu^2$, different eigenfunctions, which explicitly
depend on $\alpha_s(\mu^2)$.
Nevertheless, these eigenfunctions of the two-loop ERBL\ kernel can also
be expanded in terms of the one-loop eigenfunctions
$\Omega(x)\psi_n(x)$.
In this basis, the two-loop kernel (\ref{eq:V_ERBL_2-loop}) can be
represented in a matrix form of a triangular type:
\begin{eqnarray}
\label{eq:V_ERBL_2-loop_Matrix}
 V_\text{2-loop}(x,y;\alpha_s)
  &=& \Omega(x)
       {\sum_{n}}'{\sum_{j}}'
        \psi_n(x)\frac{V_\text{2-loop}^{n,j}(\alpha_s)}{N_j}
        \psi_j(y)\,;\\
 V_\text{2-loop}^{n,j}(\alpha_s)
  &=& \frac{-1}{2}\left(\frac{\alpha_s}{4\pi}\right)
       \left\{\gamma_0(n)\delta_{n,j}
           + \left(\frac{\alpha_s}{4\pi}\right)
              \left[\gamma_1(n)\delta_{n,j}
                   - M_{j,n}\theta(j<n)
              \right]
       \right\}
\label{eq:V_ERBL_2-loop_(n,j)}
\end{eqnarray}
with normalization coefficients $N_j$, next-to-leading order anomalous
dimensions $\gamma_1(n)$, and off-diagonal matrix elements $M_{j,n}$
given in Appendix~\ref{sec:App-A} .

We see that the diagonal terms in Eq.\ (\ref{eq:V_ERBL_2-loop_(n,j)}),
$V_\text{2-loop}^{n,n}(\alpha_s)$,
define the anomalous dimensions
\begin{eqnarray}
 \gamma_n(\alpha_s)
  &=& \frac{-1}{2}\left(\frac{\alpha_s}{4\pi}\right)
      \left[\gamma_0(n)
          + \left(\frac{\alpha_s}{4\pi}\right)\gamma_1(n)
      \right]\,,
 \label{eq:gamman}
\end{eqnarray}
whereas the off-diagonal terms,
$V_\text{2-loop}^{n,j\neq n}(\alpha_s) \sim \alpha_s^2 M_{j,n}$,
define the mixing of higher harmonics.
Solutions of the evolution equation at the two-loop level have been
given in~\cite{MR86ev,KMR86,Mul94}, and have the form
\begin{eqnarray}
 \varphi_\pi^\text{2-loop}(x,Q^2)
  = \Omega(x)\ {\sum\limits_{n}}' a_{n}(\mu_0^2)
     E_n(Q^2,\mu_0^2)
 \left[\psi_{n}(x) + \frac{\alpha_{s}(Q^2)}{4 \pi}
 {\sum\limits_{j>n}}' d_{n,j}(Q^2,\mu_0^2) \psi_{j}(x)
 \right]
  \label{eq:DA_EVO}
\end{eqnarray}
so that the evolved coefficients are
\begin{eqnarray}
 \label{eq:EVO-a_n}
 a_{n}^\text{2-loop}(Q^2)
  &=& E_n(Q^2,\mu_0^2)a_{n}(\mu_0^2)
   +  \frac{\alpha_{s}(Q^2)}{4 \pi}
       {\sum\limits_{0\leq j<n}}'
        E_j(Q^2,\mu_0^2)
         d_{j,n}(Q^2,\mu_0^2)
          a_{j}(\mu_0^2)\,.
\end{eqnarray}
Here, the ``diagonal'' part $E_n(Q^2,\mu_0^2)$ is the \emph{exact}
part of this solution, namely,
\begin{eqnarray}
 \label{eq:Exp_Evo}
   E_n(Q^2,\mu_0^2)
   &=& \frac{e_n(Q^2)}{e_n(\mu_0^2)}\,;
   \qquad
   e_n(Q^2)
   \ =\ \Big[\alpha_s(Q^2)\Big]^{\gamma(n)}
        \Big[1+\delta_1\alpha_s(Q^2)\Big]^{\omega(n)},
  \\
  \label{eq:delta1_omega}
  \delta_1
   &\equiv&
    \frac{b_1}{4\pi b_0}\,;\qquad \quad \ \
  \omega(n)\
   \equiv\
    \frac{\gamma_1(n)b_0-\gamma_0(n)b_1}
           {2b_0b_1}\,,
\end{eqnarray}
while the ``non-diagonal'' part is considered in the NLO approximation.
Note that the coefficients $d_{n,j}(Q^2,\mu_0^2)$ are related to the
off-diagonal matrix elements (\ref{eq:V_ERBL_2-loop_(n,j)}) and fix the
mixing of the higher, $j > n$, harmonics in (\ref{eq:DA_EVO}).

In the MR solution these coefficients are given by
\begin{eqnarray}
 \label{eq:MR_dnk}
 d_{j,n}^\text{MR}(Q^2,\mu_0^2)
  & = &
  \frac{M_{j,n}}{2b_0\big[\gamma(n)-\gamma(j)-1\big]}
  \left\{1
    - \left[\frac{\alpha_s(Q^2)}
                 {\alpha_s(\mu_0^2)}
      \right]^{\gamma(n)-\gamma(j)-1}
  \right\}\,.
\end{eqnarray}
As we shall see in the next sections, the approximate MR coefficients
$d_{j,n}^\text{MR}(Q^2,\mu_0^2)$
explicitly violate the group property of the RG transformation at the
$O(\alpha_s)$-level.

  \subsection{Standard formalism with modified notations}
It turns out to be more convenient to rewrite the main equations of the
previous subsection in terms of the modified notations,
$\gamma(n)$ (Eq.\ (\ref{eq:1-loop_a_n})),
$\delta_1$, and $\omega(n)$ (Eq.\ (\ref{eq:delta1_omega})).
Then, the $\beta$-function becomes
\begin{eqnarray}
 \label{eq:RG-beta_mod}
  \beta\left(\alpha_s\right)
  = -b_0\frac{\alpha_s^2}{4\pi}
      \left[1 + \delta_1 \alpha_s\right]\,.
\end{eqnarray}
Next, the two-loop kernel reads
\begin{eqnarray}
 \label{eq:V_ERBL_2-loop_(n,j)_mod}
 V_\text{2-loop}^{n,j}(\alpha_s)
  = -b_0\left(\frac{\alpha_s}{4\pi}\right)
       \left\{\left[\gamma(n)\left(1+\delta_1\alpha_s\right)
                  + \omega(n)\delta_1\alpha_s
              \right]\delta_{n,j}
            - \frac{\alpha_s}{8\pi b_0}M_{j,n}\theta(j<n)
       \right\}.
\end{eqnarray}
Finally, we rewrite the two-loop ERBL\ equations (\ref{eq:ERBL}),
(\ref{eq:V_ERBL_2-loop_Matrix}), (\ref{eq:V_ERBL_2-loop_(n,j)})
using the representation (\ref{eq:Gegenbauer}) and changing variables
from $Q^2$ to $\alpha_s$ with the help of Eqs.\ (\ref{eq:RG-coupling})
to derive
\begin{eqnarray}
  \beta(\alpha_s)\frac{da_n(\alpha_s)}{d\alpha_s}
   = \frac{-1}{2}\left(\frac{\alpha_s}{4\pi}\right)
       \left[\left(\gamma_0(n)
                 + \frac{\alpha_s}{4\pi}\gamma_1(n)
             \right)a_n(\alpha_s)
            - \frac{\alpha_s}{4\pi}
              {\sum_{0\leq j<n}}'M_{j,n}a_{j}(\alpha_s)
       \right]\! .
\end{eqnarray}
After taking into account Eqs.\ (\ref{eq:RG-beta_mod}) and
(\ref{eq:V_ERBL_2-loop_(n,j)_mod}), this expression gives
\begin{eqnarray}
 \label{eq:ERBL-2loop:a_n}
  \alpha_s\left(1+\delta_1\alpha_s\right)
   \frac{da_n(\alpha_s)}{d\alpha_s}
   = \left[\gamma(n)\left(1+\delta_1\alpha_s\right)
           + \omega(n)\delta_1\alpha_s
       \right]a_n(\alpha_s)
    - \alpha_s
       {\sum_{0\leq j<n}}'\tilde{m}_{n,j}a_{j}(\alpha_s)\,,\ \ \
\end{eqnarray}
where we have defined the reduced matrix elements
$\tilde{m}_{n,j} \equiv M_{j,n}/(8\pi b_0)$.
These expressions are going to be helpful in deriving an exact RG
solution in Sec.~\ref{sec:RG-Ex}.

\section{Group property of the renormalization group transformation for
         Gegenbauer harmonics}
 \label{sec:RGPro-2}
In this section we consider exclusively the two-loop evolution equation
given by expressions (\ref{eq:ERBL}) and (\ref{eq:V_ERBL_2-loop}).
For this reason, we omit in the following all two-loop-superscripts.
We want to understand if the approximate form (\ref{eq:DA_EVO})
respects the group property of the RG transformation, that is,
\begin{eqnarray}
 \label{eq:RG}
  U(Q^2,q^2)\cdot U(q^2,\mu^2) = U(Q^2,\mu^2)\,;
\end{eqnarray}
if not, we want to estimate to what extent it violates it.

\subsection{The second harmonic}
To this end, let us consider first the evolution of the Gegenbauer
coefficient $a_2$.
According to (\ref{eq:EVO-a_n}), we have
\begin{eqnarray}
 \label{eq:Evo-a_n_D}
 a_2(\mu_0^2) \to  a_2(Q^2)
 &\equiv& U(Q^2,\mu_0^2)a_2(\mu_0^2)\
 =\ E_2(Q^2,\mu_0^2)\, a_2(\mu_0^2)
  + D_{20}(Q^2,\mu_0^2)\, ,
\end{eqnarray}
where we defined
$D_{20}(Q^2,\mu^2)\equiv (\alpha_s(Q^2)/4\pi)d_{02}(Q^2,\mu^2)$.
If group property (\ref{eq:RG}) is valid, then it follows
\begin{eqnarray}
 E_2(Q^2,q^2)
  \big[E_2(q^2,\mu^2) a_2(\mu^2)
    +  D_{20}(q^2,\mu^2)
  \big]
 +  D_{20}(Q^2,q^2)
 \!&\!=\!&\! E_2(Q^2,\mu^2) a_2(\mu^2)
  +  D_{20}(Q^2,\mu^2)
 \nonumber
\end{eqnarray}
and from the arbitrariness of $a_2(\mu^2)$ one gets
\begin{eqnarray}
 \label{eq:RG_E2}
  E_2(Q^2,\mu^2)
 &=& E_2(Q^2,q^2)
      E_2(q^2,\mu^2)\,;\\
 \label{eq:RG_D2}
 D_{20}(Q^2,\mu^2)
 &=& E_2(Q^2,q^2)
      D_{20}(q^2,\mu^2)
   +  D_{20}(Q^2,q^2)\,.
\end{eqnarray}
The first equation is satisfied identically by virtue of
Eq.\ (\ref{eq:Exp_Evo}).
The second one is more complicated, but can be readily proved.
Define a function $Z_{20}(Q^2,\mu^2)$ by
\begin{eqnarray}
 D_{20}(Q^2,\mu^2)
 &=& e_2(Q^2)Z_{20}(Q^2,\mu^2)\,.
\end{eqnarray}
Then, Eq.\ (\ref{eq:RG_D2}) implies
\begin{eqnarray}
  \label{eq:RG_Z2}
  Z_{20}(Q^2,\mu^2)&=&Z_{20}(Q^2,q^2)+Z_{20}(q^2,\mu^2)\,.
\end{eqnarray}
The general solution of this functional equation is just
\begin{eqnarray}\label{eq:Z_2}
 Z_{20}(Q^2,\mu^2)&=&\Psi_{20}(Q^2)-\Psi_{20}(\mu^2)
\end{eqnarray}
for some auxiliary function $\Psi_{20}(\mu^2)$.
In order to find the explicit form of $\Psi_{20}(\mu^2)$, one has to
use evolution equation (\ref{eq:ERBL-2loop:a_n}), what we will
conduct in Sec.\ \ref{sec:RG-Ex}.
By using the MR approximate solution (\ref{eq:MR_dnk}), we see that
it generates the function
\begin{eqnarray}
 \label{eq:RG_D2MR}
 D_{20}^\text{MR}(Q^2,\mu^2)
 &=& \big[\alpha_s(Q^2)\big]^{\gamma(2)}
      \left[\Psi_{20}(Q^2)-\Psi_{20}(\mu^2)\right]
\end{eqnarray}
with
\begin{eqnarray}
 \label{eq:RG_FMR}
  \Psi_{20}(\mu^2)
   &=& \frac{\tilde{m}_{20}}
            {\left(\gamma(2)-1\right)
             [\alpha_s(\mu^2)]^{\gamma(2)-1}}\,.
\end{eqnarray}
From this, we conclude that in order that
$D_{20}^\text{RG}(Q^2,\mu^2)$
is consistent with the RG transformation property, it has to be given
by
\begin{eqnarray}
 D_{20}^\text{RG}(Q^2,\mu^2)&=& e_2(Q^2)
 \Big[\Psi_{20}(Q^2)-\Psi_{20}(\mu^2)\Big]\,.
 \label{eq:RG_D2Exact}
\end{eqnarray}
Comparing (\ref{eq:RG_D2MR}) with (\ref{eq:RG_D2Exact})
leads us to the conclusion that the group property of the RG
transformations in the MR approximate form is violated
because the factor $\big[1+\delta_1\alpha_s(Q^2)\big]^{\omega(2)}$
is missing.
Evidently, this violation is of $O(\alpha_s)$.

\subsection{The $n$th harmonic}
 \label{RGPro-n}
The next task is to generalize these results to the case of arbitrary
polynomial order $n=2, 4, \ldots$.
To achieve this, we define $D_{nk}(Q^2,\mu^2)$ and $Z_{nk}(Q^2,\mu^2)$
as
\begin{eqnarray}
 \label{eq:D_kn}
 D_{n,j}(Q^2,\mu^2)
 &\equiv& \frac{\alpha_s(Q^2)}{4\pi}
     E_j(Q^2,\mu^2)
      d_{j,n}(Q^2,\mu^2)
 \ =\ e_n(Q^2)Z_{n,j}(Q^2,\mu^2)e_j(\mu^2)^{-1}\,;\\
 Z_{n,j}(Q^2,\mu^2)
 &=& \frac{\alpha_s(Q^2)}{4\pi}
     e_j(Q^2)d_{j,n}(Q^2,\mu^2)e_n(Q^2)^{-1}\,.
\end{eqnarray}
Then, the evolution of the Gegenbauer coefficient $a_n$ from the
scale $\mu^2$ to the scale $Q^2$ in accordance with
Eq.\ (\ref{eq:EVO-a_n}) is given by
\begin{eqnarray}
 \label{eq:RG_a_n_}
  a_n(Q^2)
   &=& e_n(Q^2)
       \left[a_n(\mu^2)e_{n}^{-1}(\mu^2)
           + {\sum_{0\leq j<n}}'Z_{n,j}(Q^2,\mu^2)\,
              a_{j}(\mu^2)e_{j}^{-1}(\mu^2)
       \right]\,.
\end{eqnarray}
The group property (\ref{eq:RG}) dictates
\begin{eqnarray}
 \!\!\!\!
 {\sum_{0\leq j<n}}'Z_{n,j}(Q^2,\mu^2)\, a_{j}(\mu^2)e_{j}(\mu^2)^{-1}
 = {\sum_{0\leq j<n}}'
      \left[Z_{n,j}(Q^2,q^2)+Z_{n,j}(q^2,\mu^2)
      \right]\, a_{j}(\mu^2)e_{j}^{-1}(\mu^2)\nonumber\\
 ~~~~~~~~~~~~~~~~~~~
 + {\sum_{0\leq j<n}}'{\sum_{0\leq k<j}}'
              Z_{n,j}(Q^2,q^2)Z_{jk}(q^2,\mu^2)\,
               a_{k}(\mu^2)e_{k}^{-1}(\mu^2)
\end{eqnarray}
and from the arbitrariness of $a_{j}(\mu^2)$ we obtain
\begin{eqnarray}
 \label{eq:RG_Znj}
 Z_{n,j}(Q^2,\mu^2)
 &=& Z_{n,j}(Q^2,q^2) + Z_{n,j}(q^2,\mu^2)
  + {\sum_{n>k>j}}'Z_{nk}(Q^2,q^2)Z_{kj}(q^2,\mu^2)\,.
\end{eqnarray}

\subsubsection{The case $j=n-2$}
For $j=n-2$ we have the complete analogue of Eq.\ (\ref{eq:RG_Z2}),
i.e.,
\begin{eqnarray}
 \label{eq:Z_n,n-2}
 Z_{n,n-2}(Q^2,\mu^2)
 &=& Z_{n,n-2}(Q^2,q^2) + Z_{n,n-2}(q^2,\mu^2)
\end{eqnarray}
yielding the exact solution
\begin{eqnarray}
\label{eq:GenSol_Znn-2}
  Z_{n,n-2}(Q^2,\mu^2)&=&\Psi_{n,n-2}(Q^2)-\Psi_{n,n-2}(\mu^2)\,.
\end{eqnarray}

\subsubsection{The case $j=n-k$ with arbitrary and even $k<n$}
\label{sec:case of j=n-k}
Let us rewrite Eq.\ (\ref{eq:RG_Znj}) in the more appropriate form
\begin{eqnarray}
 \label{eq:RG_Zn,n-k}
 Z_{n,n-k}(Q^2,\mu^2)
 &=& Z_{n,n-k}(Q^2,q^2) + Z_{n,n-k}(q^2,\mu^2)
 \nonumber\\
 & +& {\sum_{0<j<k}}'Z_{n,n-j}(Q^2,q^2)Z_{n-j,n-k}(q^2,\mu^2)\,.
\end{eqnarray}
Using the principle of mathematical induction, we can prove that
\begin{eqnarray}
 \label{eq:sol_Zn,n-k}
 Z_{n,n-k}(Q^2,\mu^2) = \Psi_{n,n-k}(Q^2)-\Psi_{n,n-k}(\mu^2)
 - {\sum_{0<j<k}}'Z_{n,n-j}(Q^2,\mu^2)\Psi_{n-j,n-k}(\mu^2)
\end{eqnarray}
is the solution of Eq.\ (\ref{eq:RG_Zn,n-k})
(relegating further details to Appendix~\ref{sec:App-MI}).

We are now in the position to summarize our findings and rewrite the
solution of the NLO evolution equation (\ref{eq:ERBL}) in the form
(\ref{eq:Gegenbauer}) with the $n$th Gegenbauer coefficient that
respects the RG properties being given by
\begin{eqnarray}
\label{eq:RG_a_n}
 a_n^\text{RG}(\mu^2)
  &=& e_n(\mu^2)
      \left[a_n(\mu_0^2)e_{n}^{-1}(\mu_0^2)
          + {\sum_{0\leq j<n}}'Z_{n,j}^\text{RG}(\mu^2,\mu_0^2)\,
            a_{j}(\mu_0^2)e_{j}^{-1}(\mu_0^2)
      \right]\,,
\end{eqnarray}
where
\begin{eqnarray}
 e_n(\mu^2)
  &=& \Big[\alpha_s(\mu^2)\Big]^{\gamma(n)}
       \Big[1+\delta_1\alpha_s(\mu^2)\Big]^{\omega(n)}\,;\\
\label{eq:RG_Z_n,k}
 Z_{n,k}^\text{RG}(\mu^2,\mu_0^2)
  &=& \Psi_{n,k}(\mu^2)-\Psi_{n,k}(\mu_0^2)
   - {\sum_{n>j>k}}'Z^\text{RG}_{n,j}(\mu^2,\mu_0^2)
     \Psi_{j,k}(\mu_0^2)\, ,
\end{eqnarray}
with $\Psi_{n,j}(\mu^2)$ being defined on account of evolution equation
(\ref{eq:ERBL-2loop:a_n}), a task we will conduct in Sec.
\ref{sec:RG-Ex}.

It is instructive to rewrite the approximate MR solution in an
analogous manner to obtain
\begin{eqnarray}
 a_n^\text{MR}(\mu^2)
  &=& e_n(\mu^2)
      \left[a_{n}(\mu_0^2)e_n^{-1}(\mu_0^2)
         +  {\sum_{0\leq j<n}}'Z_{n,j}^\text{MR}(\mu^2,\mu_0^2)
            a_{j}(\mu_0^2)e_{j}^{-1}(\mu_0^2)
       \right]\,;\\
\label{eq:MR_Zn,n-k}
 Z_{n,k}^\text{MR}(\mu^2,\mu_0^2)
 &\equiv&
    \frac{e_{k}(\mu^2)\alpha_s(\mu^2)^{\gamma(n)}}
         {e_{n}(\mu^2)\alpha_s(\mu^2)^{\gamma(k)}}
    \left[\Psi_{n,k}^\text{MR}(\mu^2)
        - \Psi_{n,k}^\text{MR}(\mu_0^2)
    \right]\nonumber\\
 &=&\left[1+\delta_1\alpha_s(\mu^2)\right]^{\omega(k)-\omega(n)}
    \left[\Psi_{n,k}^\text{MR}(\mu^2)
        - \Psi_{n,k}^\text{MR}(\mu_0^2)
    \right]\,,\\
 \label{eq:RG_Psi_nn-k}
  \Psi_{n,n-k}^\text{MR}(Q^2)
 &=&
    \frac{\tilde{m}_{n,n-k}}
         {\left[\gamma(n)-\gamma(n-k)-1\right]
          \left[\alpha_s(Q^2)\right]^{\gamma(n)-\gamma(n-k)-1}}\,.
\end{eqnarray}
Comparing Eqs.\ (\ref{eq:RG_Z_n,k}) and (\ref{eq:MR_Zn,n-k}), we see
that $Z_{n,k}^\text{RG}(\mu^2,\mu_0^2)$
differs from
$ Z_{n,k}^\text{MR}(\mu^2,\mu_0^2)$ in two respects:
\begin{itemize}
  \item It contains a factor
        $\left[1+\delta_1\alpha_s(\mu^2)
         \right]^{\omega(k)-\omega(n)}$
        in the leading term $\Psi_{n,k}(\mu^2)-\Psi_{n,k}(\mu_0^2)$.
  \item It comprises additional terms of the type
        $Z_{n,j}(\mu^2,\mu_0^2)\Psi_{j,k}(\mu_0^2)$,
        which are related to $Z_{n,j}$ bearing a smaller index $j<k$.
\end{itemize}

\section{Exact renormalization group solution}
 \label{sec:RG-Ex}
\begin{figure}[t]
 \centerline{\includegraphics[width=\textwidth]{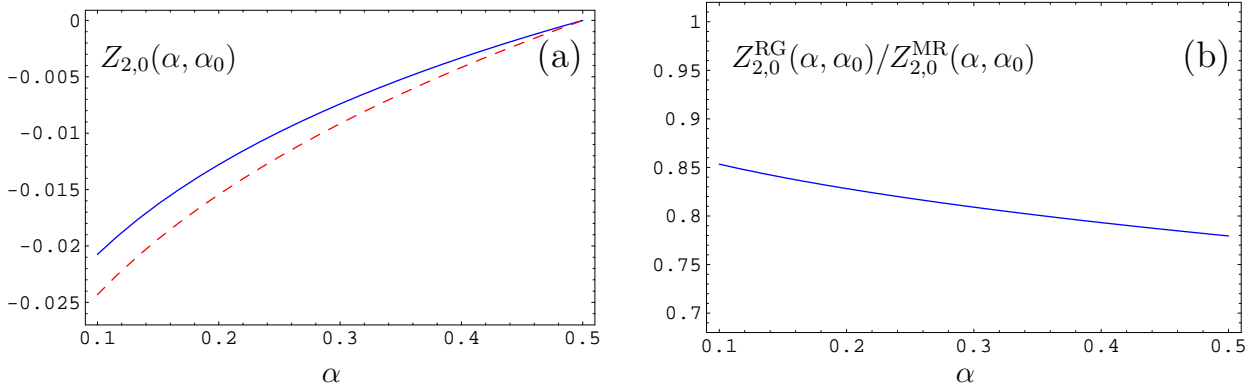}}%
  \caption{\label{fig:comp_Z20}\footnotesize
    (a) The solid line corresponds to
    $Z_{2,0}^\text{RG}(\alpha,\alpha_0)$ (RG-improved case)
    and the dashed line to
    $Z_{2,0}^\text{MR}(\alpha,\alpha_0)$ (MR-approximate case).
    (b) The ratio
    $Z_{2,0}^\text{RG}(\alpha,\alpha_0)/
    Z_{2,0}^\text{MR}(\alpha,\alpha_0)$ is plotted vs.\ $\alpha$.
    In both panels we use $\alpha_0=0.5$.}
\end{figure}

Up to this point we have presented only an approximate form of the
$\Psi$-functions (see, Eq.\ (\ref{eq:RG_Psi_nn-k})), derived from the
MR solution, that satisfy the RG equation.
Now that we have outlined the basic steps of the RG restoration, we
proceed to our analysis of its consequences for the
$\Psi_{n,k}(\mu^2)$-functions.
Substituting our solution (\ref{eq:RG_Z_n,k}) into the two-loop ERBL\
equation (\ref{eq:ERBL-2loop:a_n}), we obtain\footnote{%
From now on we trade all the $\mu^2$-dependence for an
$\alpha_s$-dependence (with $\alpha_0=\alpha_s(\mu_0^2)$).}
\begin{eqnarray}
 \left(1+\delta_1\alpha_s\right)
   e_n(\alpha_s)
   {\sum_{0\leq j<n}}'
    \frac{dZ_{n,j}^\text{RG}(\alpha_s,\alpha_0)}{d\alpha_s}\,
     a_{j}(\alpha_0)\,e_{j}^{-1}(\alpha_0)\
    =\qquad\qquad\qquad\qquad\quad\nonumber\\
    =\ - {\sum_{0\leq j<n}}'
          \frac{M_{j,n}}{8\pi b_0}\,
           e_{j}(\alpha_s)
     \left[a_j(\alpha_0)e_j^{-1}(\alpha_0)
           + {\sum_{0\leq k<j}}' Z_{j,k}^\text{RG}(\alpha_s,\alpha_0)\,
             a_k(\alpha_0)e_k^{-1}(\alpha_0)
         \right]\,.
\end{eqnarray}
Changing indices of summation in the double sum from $j,k$  to $k,j$
and rearranging them as to sum first over $j$, we obtain due to the
arbitrariness of $a_j(\alpha_0)$:
\begin{eqnarray}
 \label{eq:ERBL-2loop:Z_nj}
  \left(1+\delta_1\alpha_s\right)e_n(\alpha_s)
   \frac{dZ_{n,j}^\text{RG}(\alpha_s,\alpha_0)}{d\alpha_s}
   = - \tilde{m}_{n,j}\,e_{j}(\alpha_s)
     - {\sum_{j<k<n}}'\tilde{m}_{n,k}\,
        Z_{k,j}^\text{RG}(\alpha_s,\alpha_0)
         e_{k}(\alpha_s)\,.
\end{eqnarray}
For $j=n-2$ this differential equation reduces to
\begin{eqnarray}
 \label{eq:dZ/da_nn-2}
  - \frac{dZ_{n,n-2}^\text{RG}(\alpha,\alpha_0)}{d\alpha}
   &=& \tilde{m}_{n,n-2}
        \frac{e_{n-2}(\alpha)}
             {e_n(\alpha)\left(1+\delta_1\alpha\right)}\,,
\end{eqnarray}
whose exact solution is
\begin{eqnarray}
 \label{eq:Exact_Z_n,n-2}
  Z_{n,n-2}^\text{RG}(\alpha,\alpha_0)
   = \Psi_{n,n-2}^\text{RG}(\alpha)
   - \Psi_{n,n-2}^\text{RG}(\alpha_0)
\end{eqnarray}
with
\begin{subequations}
\begin{eqnarray}
 \label{eq:RGimp_Psi_nn-2}
  \Psi_{n,n-2}^\text{RG}(\alpha)
  &=& \Phi_{n,n-2}^\text{RG}\left(\alpha\right)\,;  \\
 \Phi_{n,j}^\text{RG}\left(\alpha\right)
  &\equiv&
   \tilde{m}_{n,j}\,
    \frac{_{2}F_{1}\left(1-\gamma(n)+\gamma(j),
                         1+\omega(n)-\omega(j),
                         2-\gamma(n)+\gamma(j),
                         -\delta_1\alpha\right)}
       {\left[\gamma(n)-\gamma(j)-1\right]
        \alpha^{\gamma(n)-\gamma(j)-1}}\,.
 ~~~\label{eq:RGimp_Phi_nk}
\end{eqnarray}
\end{subequations}
By definition, the function $\Phi_{n,k}^\text{RG}(\alpha)$ satisfies
the following evolution equation:
\begin{eqnarray}
 \label{eq:dPhi/da_nk}
  \frac{d\Phi_{n,j}^\text{RG}(\alpha)}{d\alpha}
   &=& -\ \frac{\tilde{m}_{n,j}\ e_{j}(\alpha)}
            {e_n(\alpha)\left(1+\delta_1\alpha\right)}\,.
\end{eqnarray}
In Fig.~\ref{fig:comp_Z20} we show both solutions,
 $Z_{2,0}^\text{RG}(\alpha,\alpha_0)$ and
 $Z_{2,0}^\text{MR}(\alpha,\alpha_0)$,
for $\alpha<\alpha_0=0.5$.
We observe that the error in $Z_{2,0}(\alpha,\alpha_0)$, when using
the approximate MR solution, varies from 25\%
($\alpha\approx\alpha_0$) to 15\% ($\alpha\to0.1$).

Consider now Eq.\ (\ref{eq:ERBL-2loop:Z_nj}) in the general case
$k=n-j$ with $2\leq j<n$.
We have the RG representation of its solution (\ref{eq:RG_Z_n,k}),
which can be substituted into the evolution equation
(\ref{eq:ERBL-2loop:Z_nj}) to obtain\footnote{%
We use here the same trick of interchanging summation indices
$(j,k)\to (k,j)$ and performing the sums in $j$ and $k$,
as in Appendix \ref{sec:App-MI}.}
\begin{eqnarray}
 \label{eq:ERBL-2loop:Psi_nj}
  \frac{d\Psi_{n,j}(\alpha)}{d\alpha}
   = - \tilde{m}_{n,j}
        \frac{e_{j}(\alpha)}{e_n(\alpha)\left(1+\delta_1\alpha\right)}
     - {\sum_{j<k<n}}'
       \tilde{m}_{n,k}
        \frac{\Psi_{k,j}(\alpha)e_{k}(\alpha)}
             {e_n(\alpha)\left(1+\delta_1\alpha\right)}\,.
\end{eqnarray}
The solution of this equation can be represented in terms of
quadratures:
\begin{eqnarray}
 \label{eq:2loop_Psi_nj}
  \Psi_{n,j}^\text{RG}(\alpha)
   = \Phi_{n,j}^\text{RG}(\alpha)
     - {\sum_{j<k<n}}'
       \tilde{m}_{n,k}
        \int_{0}^{\alpha}
         \frac{\Psi^\text{RG}_{k,j}(a)e_{k}(a)}
              {e_n(a)\left(1+\delta_1a\right)}d\,a\,.
\end{eqnarray}
Before analyzing this solution, let us sketch the procedure to derive
it.
\begin{itemize}
  \item We analyzed the NLO evolution equations
        (\ref{eq:ERBL-2loop:a_n}).
  \item We used for $a_n(\mu^2)$ the representation (\ref{eq:RG_a_n})
        in terms of $Z^\text{RG}_{n,j}(\mu^2,\mu_0^2)$ and obtained
        Eq.\ (\ref{eq:ERBL-2loop:Z_nj}).
  \item We employed the RG representation (\ref{eq:RG_Z_n,k}) for
        $Z^\text{RG}_{n,j}(\mu^2,\mu_0^2)$ and derived this way
        Eq.\ (\ref{eq:ERBL-2loop:Psi_nj}) which yields for
        $\Psi_{n,j}^\text{RG}(\alpha)$ an explicit solution given by
        expression (\ref{eq:2loop_Psi_nj}).
\end{itemize}

An exact solution of this equation for the case $k=n-2$ is provided by
$\Phi_{n,n-2}^\text{RG}(\alpha)$, (\ref{eq:RGimp_Phi_nk}), expressed
in terms of the hypergeometric function
$_{2}F_{1}\left(a,b,1+a,-\delta_1\alpha\right)$
with
$a=1-\gamma(n)+\gamma(n-2)$ and
$b=1+\omega(n)-\omega(n-2)$.
From Eq. (\ref{eq:2loop_Psi_nj}) we can restore the leading asymptotics
of $\Psi_{n,j}^\text{RG}(\alpha)$ to get
\begin{eqnarray}
 \label{eq:Psi_nj_Asymptotics}
  \Psi_{n,j}^\text{RG}(\alpha)
   &\stackrel{\alpha\to0}{\rightarrow}&
    \alpha^{1+\gamma(j)-\gamma(n)}\,,
\end{eqnarray}
a form dictated by the $\Phi_{n,j}^\text{RG}(\alpha)$-term,
whereas the term containing the ${\sum}'$ in Eq.\
(\ref{eq:2loop_Psi_nj}) generates a correction with the asymptotics
$\alpha^{2+\gamma(j)-\gamma(n)}$.
This property allows us to suggest an approximate RG solution of the
NLO evolution equation (indicated below by the superscript `app');
viz.,
\begin{figure}[h]
 \centerline{\includegraphics[width=0.48\textwidth]{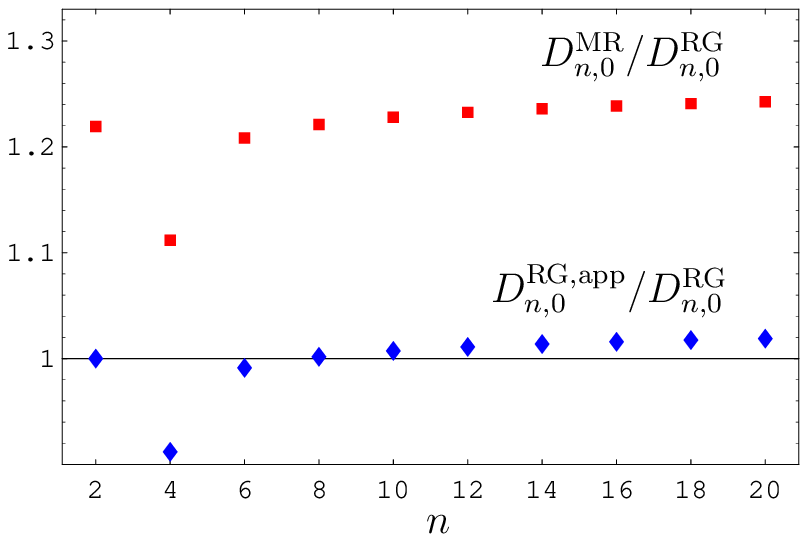}%
  ~~~\includegraphics[width=0.48\textwidth]{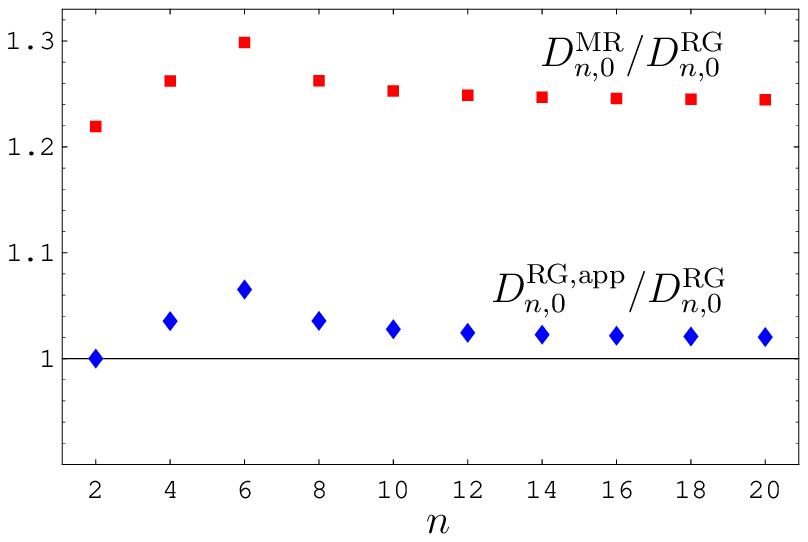}}%
  \caption{\label{fig:comp_Ev_App}\footnotesize
    Comparison of the RG improved, RG approximated, and original MR
    evolution in terms of the coefficients
    $D_{n,0}(\alpha_1,\alpha_0)$
    with
    $\alpha_1=\alpha_s(4~\gev{2})$ and $\alpha_0=\alpha_s(1~\gev{2})$.
    The blue points at the bottom denote
    $D_{n,0}^\text{RG}(\alpha_1,\alpha_0)/
    D_{n,0}^\text{MR}(\alpha_1,\alpha_0)$,
    whereas the red points above them represent
    $D_{n,0}^\text{RG}(\alpha_1,\alpha_0)/
    D_{n,0}^\text{RG,app}(\alpha_1,\alpha_0)$.
    Left panel: The original values of $M_{k,n}$ with $M_{0,4}=0.285$
    are used.
    Right panel: All coefficients $M_{k,n}$ have the original values,
    except for $M_{0,4}$ for which we set $M_{0,4}=-0.8$.}
\end{figure}
\begin{eqnarray}
 \label{eq:Appro_Psi_nk}
  \Psi_{n,k}^\text{RG,app}(\alpha)
   &=& \Phi_{n,k}^\text{RG}(\alpha)\,;
\\ \label{eq:Appro_Z_nk}
  Z_{n,k}^\text{RG,app}(\alpha,\alpha_0)
   &=& \Phi_{n,k}^\text{RG}(\alpha)
   - \Phi_{n,k}^\text{RG}(\alpha_0)\,.
\end{eqnarray}
The results for the coefficients $D_{n,0}(\alpha_1,\alpha_0)$ are shown
in Fig.\ \ref{fig:comp_Ev_App}.
As already mentioned, the RG-approximate evolution is rather good for
$n\geq6$, but for $n=4$ it drops to a much too low value compared to
the exact RG-improved result.
The reason for this ``jump'' can be traced to the (unexpected)
smallness and positiveness of $M_{0,4} = 0.285$, relative to
$M_{0,2} = -6.01$ and $M_{2,4} = -17.05$, cf.\ Eq.\ (\ref{eq:M024}).
To show that this jump
is connected to the particular value of $M_{0,n} = 0.285$
at  $n=4$, and is not a numerical error,
we show in the right panel of Fig.\ \ref{fig:comp_Ev_App}
a graphics,
where we set by hand the value of  $M_{0,n} = -0.8$.
As one sees, the result of this rough simulation
yields to a complete cancellation of the dip at $n=4$
turning it into a slight bump at $n=6$,
proving the consistency of our numerical algorithm.

The numerical solution of Eq.\ (\ref{eq:2loop_Psi_nj}) proceeds through
the following step-by-step procedure:
\begin{itemize}
  \item Given that we know the solution for $j=n-2$---supplied by
  $\Psi_{n,n-2}^\text{RG}(\alpha)=\Phi_{n,n-2}^\text{RG}(\alpha)$,
  cf.\ Eq.\ (\ref{eq:RGimp_Phi_nk})---we determine
  $\Psi_{n-2,n-4}^\text{RG}(\alpha)$ and solve numerically
  Eq.\ (\ref{eq:2loop_Psi_nj}) for $j=n-4$.
  This yields
  \begin{eqnarray}
    \Psi_{n,n-4}^\text{RG}(\alpha)
   = \Phi_{n,n-4}^\text{RG}(\alpha)
     - \tilde{m}_{n,n-2}
        \int_{\alpha_{\infty}}^{\alpha}
        \frac{\Psi^\text{RG}_{n-2,n-4}(a)e_{n-2}(a)}
             {e_n(a)\left(1+\delta_1a\right)}d\,a\,.
     \nonumber
  \end{eqnarray}
  \item We then solve numerically Eq.\ (\ref{eq:2loop_Psi_nj}) for
  $j=n-k$ using the results
  $\{\Psi_{n-k+2,n-k}^\text{RG}(\alpha), \ldots,
     \Psi_{n-2,n-k}^\text{RG}(\alpha)\}$
  of the previous steps to get $\Psi_{n,n-k}^\text{RG}(\alpha)$.
  \item Finally, we solve numerically Eq.\ (\ref{eq:2loop_Psi_nj})
  for $j=0$, employing
  $\{\Psi_{2,0}^\text{RG}(\alpha), \ldots,
     \Psi_{n-2,0}^\text{RG}(\alpha)\}$
  to find $\Psi_{n,0}^\text{RG}(\alpha)$.
\end{itemize}
This procedure is repeated for all values of $n$, starting with $n=2$,
continuing with $n=4$, and so on.
\begin{figure}[t]
 \centerline{\includegraphics[width=\textwidth]{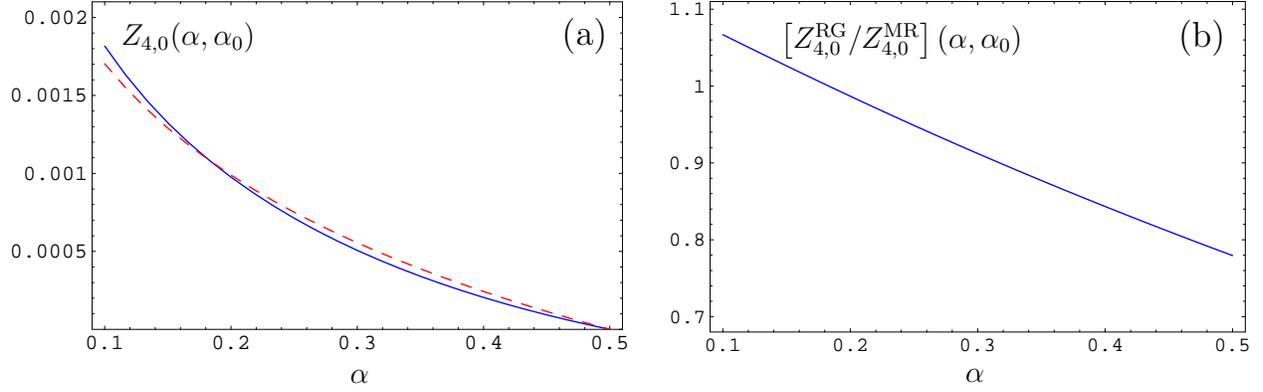}}%
  \caption{\label{fig:comp_Z40}\footnotesize
    (a) The solid line corresponds to
    $Z_{4,0}^\text{RG}(\alpha,\alpha_0)$ and the dashed line to
    $Z_{4,0}^\text{MR}(\alpha,\alpha_0)$.
    (b) The ratio $Z_{4,0}^\text{RG}(\alpha,\alpha_0)/
    Z_{4,0}^\text{MR}(\alpha,\alpha_0)$ vs. $\alpha$ is shown.
    We use in both panels $\alpha_0=0.5$.}
\end{figure}
The upshot of this procedure, $Z_{4,0}^\text{RG}(\alpha,\alpha_0)$, is
shown in Fig.\ ~\ref{fig:comp_Z40} in comparison with the approximate
MR solution  $Z_{4,0}^\text{MR}(\alpha,\alpha_0)$, using in both cases
$\alpha<\alpha_0=0.5$.
One notes that the error in $Z_{4,0}(\alpha,\alpha_0)$
induced by the MR approximation
is quite substantial,
varying from $+20$\% ($\alpha\approx\alpha_0$) to
$-5$\% ($\alpha\to0$).

\section{Numerical importance of the renormalization group solution}
 \label{sec:Numerical}
\begin{figure}[b]
\centerline{\includegraphics[width=0.9\textwidth]{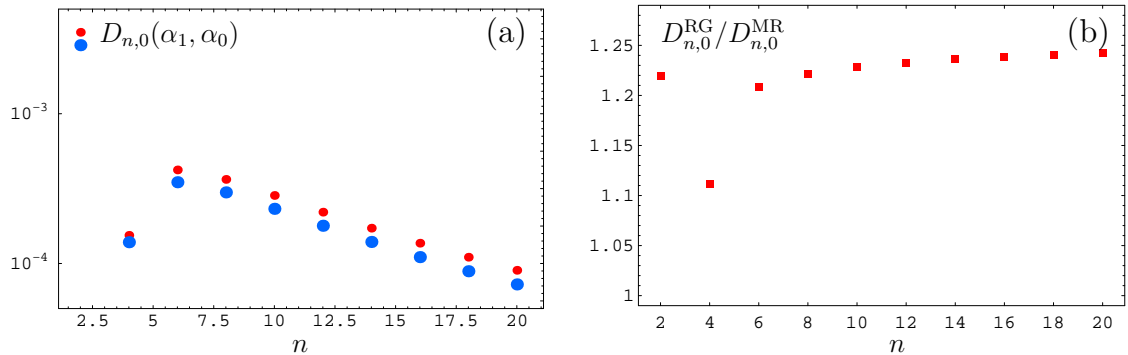}}%
\caption{\label{fig:comp_Ev}\footnotesize
    Comparison of the RG-improved evolution coefficients
    $D_{n,0}(\alpha_1,\alpha_0)$ with $\alpha_1=\alpha_s(4~\gev{2})$
    and
    $\alpha_0=\alpha_s(1~\gev{2})$ contrasted to the MR-approximate
    version.
    (a) The blue points denote
    $D_{n,0}^\text{RG}(\alpha_1,\alpha_0)$,
    whereas the red points mark $D_{n,0}^\text{MR}(\alpha_1,\alpha_0)$
    (note the logarithmic scale used for the ordinate.)
    (b) The ratio $D_{n,0}^\text{RG}(\alpha_1,\alpha_0)/
    D_{n,0}^\text{MR}(\alpha_1,\alpha_0)$ as a function of the order
    number $n$ is shown.}
\end{figure}

Let us now outline the importance of the RG improvement in terms of
the evolution of the one-loop asymptotic DA using the MR scheme and
the RG-improved one in comparison
and continue then to the more complicated case
of a double-humped, endpoint-suppressed pion DA
obtained in~\cite{BMS01}.
The asymptotic solution to the one-loop evolution equation reads
\begin{eqnarray}
 \label{eq:input}
  \varphi_\pi(x,\mu_0^2=1~\gev{2})
   &=& \varphi_\text{as}(x)
   \ =\ 6 x (1-x)\,;\qquad
   a_n(\mu_0^2) = \Big\{\begin{array}{lcl}
    1 &~& \text{if~}n=0 \\
    0 &~& \text{if~}n\neq0\end{array}\,.
\end{eqnarray}
Then, in accordance with (\ref{eq:RG_a_n_}), we have
\begin{eqnarray}
  a_n(Q^2)
   &=& \Big\{\begin{array}{lcl}
    1 &~& \text{if~}n=0 \\
    e_n(Q^2)Z_{n,0}(Q^2,\mu_0^2) = D_{n,0}(Q^2,\mu_0^2)
&~& \text{if~}n\neq0\end{array}.
\end{eqnarray}
Figure \ref{fig:comp_Ev}(a) shows the comparison of the two evolution
schemes for the coefficients $D_{n,0}(Q^2,\mu_0^2)$ for $Q^2=4$~GeV$^2$
with $n=2,\ldots,20$, whereas the ratio of these coefficients is
displayed in Fig.~\ref{fig:comp_Ev}(b).
\begin{figure}[b]
 \centerline{\includegraphics[width=0.9\textwidth]{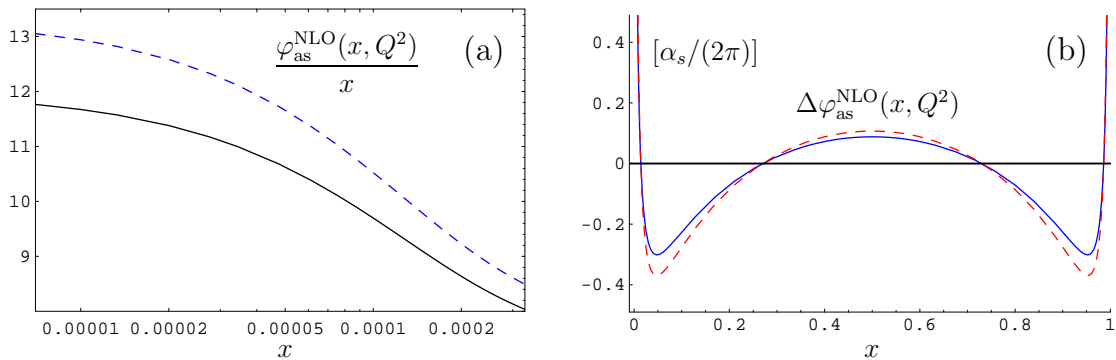}}%
  \caption{\label{fig:comp_DA}\footnotesize
   The left panel shows the extreme endpoint region
   close to $x=0$ to effect the changes entailed
   by evolution. The right panel shows in units of
   $\alpha_s/(2\pi)$ the relative NLO correction
   using the RG-corrected evolution,
   where $\Delta\varphi^\text{NLO}_\text{as}(x,Q^2)$ is defined as
   $\left[\varphi^\text{NLO}_\text{as}(x,Q^2)
   -\varphi_\text{as}(x)\right]/\varphi_\text{as}(x)$.
   In both panels, the solid line corresponds to the RG-improved
   evolution, whereas the dashed one represents the MR evolution.}
\end{figure}

Two observations are worth noting:
\begin{enumerate}
  \item[(i)] The absolute difference
   $D_{n,0}^\text{MR}(4~\gev{2},1~\gev{2})
   - D_{n,0}^\text{RG}(4~\gev{2},1~\gev{2})$
   for all $n$ is small (being of order
   $\alpha_s$---Eq.\ (\ref{eq:Evo-a_n_D})).
  \item[(ii)] The achieved improvement is high, reaching a level of
  reduction of the evolution coefficients of up to $20\%$.
\end{enumerate}

The next step of the analysis involves the comparison of the results
of the RG and MR evolution approaches
as applied to the pion DA itself---Fig.~\ref{fig:comp_DA}---taking into
account the first 100 nontrivial terms,
meaning summing up $n=2,4,\ldots 200$ higher Gegenbauer harmonics.
In panel (a) we display the function $\varphi(x)/x$ in the vicinity of
the endpoint $x=0$, whereas in panel (b) we show the DA around the
middle point $x=0.5$.
The key observation here is that in both panels, (a) and (b), the
RG-improved approach produces slightly smaller results.
More precisely, the DA gradient for the DA value turns out to be
smaller by 4\% at the origin and by 0.1\% at the middle point $x=0.5$.

In our recent papers with S.~V.~Mikhailov \cite{BMS02}, dealing with the
extraction of constraints on the Gegenbauer coefficients of the pion DA
from the CLEO data \cite{CLEO98} on the $\gamma^*\gamma\to\pi$
transition form factor
$F_{\gamma^*\gamma\pi^0}(Q^2)$,
we used the standard MR approach to evolve the pion DA~\cite{BMS01}
from the normalization scale $\mu_{0}^{2}\simeq 1$~GeV${}^{2}$
to the  scale $\mu_\text{SY}^2=5.76$ GeV$^2$, introduced in \cite{SY99}
by Schmedding and Yakovlev (SY), as being relevant for the CLEO
experiment.
These results are compiled in Table~\ref{tab:BMS_a24}.
One appreciates that the evolution improvement here due to the RG
modification is of minor importance, reaching just the order of 1\%.
\begin{table}[h]
\caption{Results of the evolution of the Gegenbauer coefficients $a_2$
         and $a_4$ from the scale $\mu_0^2=1$~GeV$^2$ to the scale
         $\mu_\text{SY}^2=5.76$ GeV$^2$~\cite{SY99} for the asymptotic,
         BMS \cite{BMS01}, and CZ~\cite{CZ84} pion DAs.
         \label{tab:BMS_a24}}
\begin{ruledtabular}
\begin{tabular}{ccccccc}
DA models
     & $a_2(\mu_0^2)$
        & $a_2^\text{MR}(\mu_\text{SY}^2)$
            & $a_2^\text{RG}(\mu_\text{SY}^2)$
                & $a_4(\mu_0^2)$
                    & $a_4^\text{MR}(\mu_\text{SY}^2)$
                        & $a_4^\text{RG}(\mu_\text{SY}^2)$
                   \\ \hline \hline
As   & 0
        & $-0.004$
            & $-0.003$
                & 0
                    & $0$
                        & $0$
             \\
BMS  & $0.204$
        & $0.144$
            & $0.145$
                & $-0.144$
                    & $-0.093$
                        & $-0.092$
              \\
CZ  & $0.56$
        & $0.403$
            & $0.403$
                & 0
                    & -$0.004$
                        & -$0.005$
              \\
\end{tabular}
\end{ruledtabular}
\end{table}

\begin{figure}[b]
 \centerline{\includegraphics[width=0.48\textwidth]{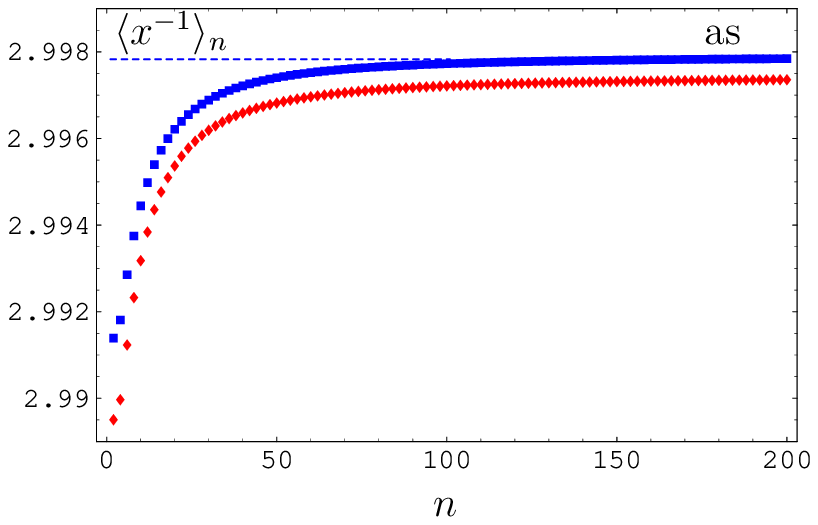}~~~
 \includegraphics[width=0.48\textwidth]{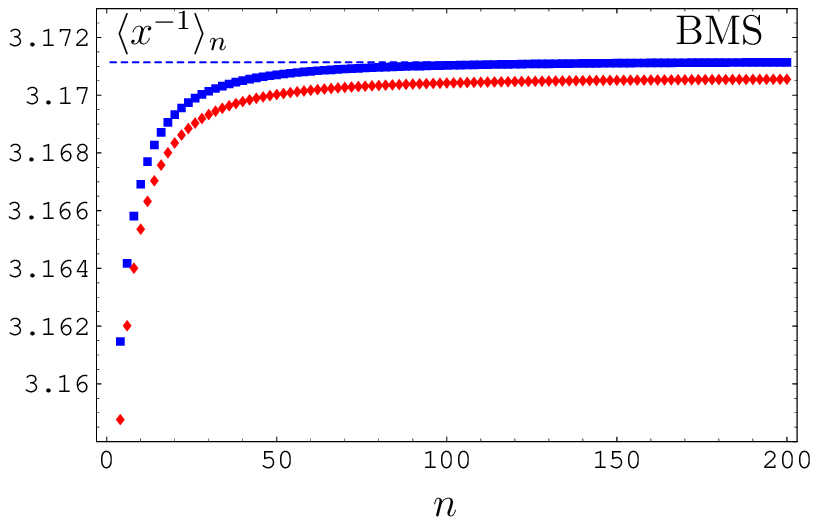}}%
  \caption{\label{fig:comp_IM}\footnotesize
    The impact on the asymptotic pion DA and the BMS one
    (indicated by corresponding acronyms)
    of the RG-improved evolution (blue squares on the top)
    relative to the MR evolution (red diamonds)
    on the truncated inverse moment
    $\va{x^{-1}}_n$,  Eq.\ (\protect{\ref{eq:inv-mom}}),
    as a function of the number $n$ of the Gegenbauer harmonics
    included.
    The limiting value $\va{x^{-1}}_\infty^\text{RG}$
    (dashed line) is also displayed. }
\end{figure}

In Fig.~\ref{fig:comp_IM}(a) we show the same sort of comparison for
the truncated inverse moment:
\begin{eqnarray}\label{eq:inv-mom}
  \va{x^{-1}}_n &=& 3\sum_{j\leq n}a_j(Q^2)\,.
\end{eqnarray}
We see that in the RG-improved evolution scheme, $\va{x^{-1}}_n$
approaches the limiting value $\va{x^{-1}}_\infty$,
represented by the dashed line,
more rapidly.

Note in this context that in a previous analysis with
K.~Passek-Kumeri\v{c}ki and  W.~Schroers of the pion form factor
\cite{BPSS04} we have analyzed the NLO evolution of the pion DA in
the MR evolution scheme, using the property of the two-loop evolution
just described.
More precisely, if one is interested only in the value of the inverse
moment, as this is exactly the case for the factorized part of the
pion form factor, then it is actually enough to use for the
calculation the LO evolution (\ref{eq:1-loop_a_n}), ensuring this
way an accuracy at the 1\% level.
Indeed, to establish this property, we analyzed in~\cite{BPSS04}
numerically the convergence of the truncated moment $\va{x^{-1}}_n$
up to $n=100$.
In the present RG-improved approach,
this property can be established even at lower values of $n$.

Let us step one level higher and consider now the BMS pion DA
\begin{eqnarray}
 \label{eq:inputBMS}
 \varphi_\text{BMS}(x)
   \ =\ 6 x (1-x)\,
   \left[1
   + a_2^\text{BMS}(\mu_0^2)\,C^{3/2}_2(2x-1)
   + a_4^\text{BMS}(\mu_0^2)\,C^{3/2}_4(2x-1)
   \right]
\end{eqnarray}
as the initial input of evolution.
Then, in accordance with (\ref{eq:RG_a_n_}), we have
\begin{eqnarray}
 \label{eq:a2BMS}
  a_2^\text{BMS}(Q^2)
   &=& E_{2}(Q^2,\mu_0^2)\,a_2^\text{BMS}(\mu_0^2)
     + D_{2,0}(Q^2,\mu_0^2)\,;\\
 \label{eq:a4BMS}
  a_4^\text{BMS}(Q^2)
   &=& E_{4}(Q^2,\mu_0^2)\,a_4^\text{BMS}(\mu_0^2)
     + D_{4,0}(Q^2,\mu_0^2)
     + D_{4,2}(Q^2,\mu_0^2)\,a_2^\text{BMS}(\mu_0^2)\,.
\end{eqnarray}
All higher coefficients at the initial point $\mu_0^2=1$~GeV$^2$
vanish
and therefore we have
\begin{eqnarray}
 \label{eq:a6BMS}
  a_{n\geq6}^\text{BMS}(Q^2)
   &=& D_{n,0}(Q^2,\mu_0^2)
     + D_{n,2}(Q^2,\mu_0^2)\,a_2^\text{BMS}(\mu_0^2)
     + D_{n,4}(Q^2,\mu_0^2)\,a_4^\text{BMS}(\mu_0^2)\,.
\end{eqnarray}
Then, in NLO, the BMS pion DA
evolved from the scale $\mu_0^2=1$~GeV$^2$ to the scale $Q^2$,
within the RG-improved scheme,
is given by
\begin{eqnarray}
 \label{eq:evolvedBMS}
 \varphi_\text{BMS}^\text{NLO}(x,Q^2)
   \ =\ 6 x (1-x)\,
   \left[1
   + \sum_{n\geq1}a_{2n}^\text{BMS}(Q^2)\,C^{3/2}_{2n}(2x-1)
   \right]\,,
\end{eqnarray}
whereas its counterpart in LO,
$\varphi_\text{BMS}^\text{LO}(x)$,
is obtained with the diagonal part
given by Eqs.\ (\ref{eq:a2BMS})--(\ref{eq:a4BMS}).

\begin{figure}[t]
 \centerline{\includegraphics[width=0.9\textwidth]{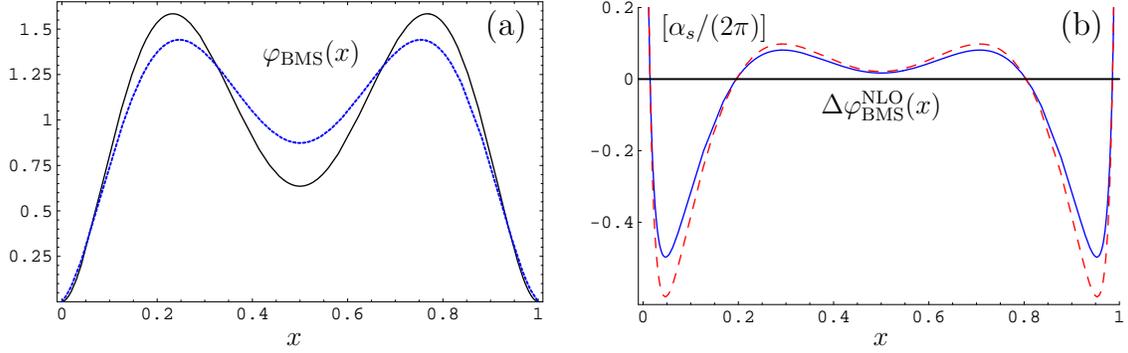}}%
  \caption{\label{fig:comp_BMS}\footnotesize
   (a) The effect of the RG-improved evolution (blue dashed line)
   on the initial BMS pion DA (solid line) is illustrated.
   (b) We show in units of $\alpha_s/(2\pi)$ the relative NLO
   correction using on the initial BMS pion DA the RG-improved
   evolution, the emphasis being placed on the endpoint regions $x=0$
   and $x=1$.
   Notice that $\Delta\varphi_\text{BMS}^\text{NLO}(x)$ is defined as
   $\left[\varphi_\text{BMS}^\text{NLO}(x,Q^2)
   -\varphi_\text{BMS}^\text{LO}(x)
   \right]/\varphi_\text{as}(x)$.
   For the sake of comparison with Fig.\ \ref{fig:comp_DA},
   we normalize this expression to $\varphi_\text{as}(x)$.}
\end{figure}

The comparison of the results
of the RG-improved and the MR-approximate evolution approaches,
as applied to the BMS pion DA,
is shown in Fig.~\ref{fig:comp_BMS}
taking into account
the first 100 nontrivial terms $(n=2,4,\ldots,200)$.
In panel (a) we display the evolution effect of the RG-improved
approach for the BMS pion DA.
For the reader's convenience,
we use in our analysis
the same numerical values
$N_f=3$ and $Q^2=4$~GeV$^2$ as in~\cite{Mul95},
but employing a higher initial point of evolution
$\mu^2=1$~GeV$^2$.
The illustration of the evolution effect,
especially on the endpoints ($x\to0$, $x\to1$),
is shown in Fig.\ \ref{fig:comp_BMS}(b).
We will return to this issue and study the influence of the
RG-improvement in our discussion of the evolution with a fixed
$\alpha_s$ in the next section.

\section{RG solution in the case of a fixed coupling constant}
 \label{sect:RG_Fixed}
\begin{figure}[t]
\centerline{\includegraphics[width=\textwidth]{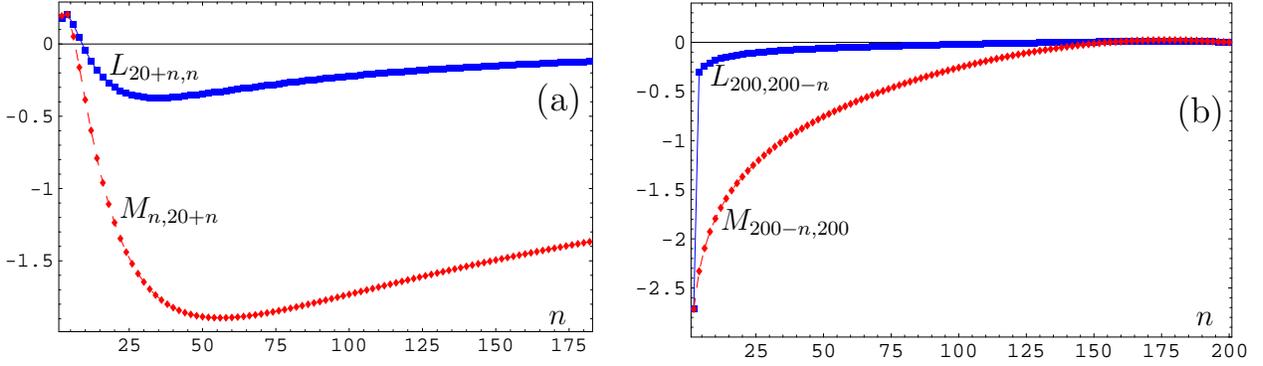}}%
\caption{\label{fig:comp_LM}\footnotesize
    Comparison of $L_{n,j}$ and $M_{j,n}$ for fixed $\alpha_s=0.5$.
    (a) The case of a fixed difference $n-j=20$ is shown.
    (b) An analogous situation is shown for $n=200$.}
\end{figure}

We turn now to a comparison of our results with those obtained in
\cite{Mul95} for the evolution with a fixed $\alpha_s$,
restricting attention to the asymptotic pion DA,
the BMS case being analogous.
To this end, let us adopt our formulae to this case,
using for the corresponding quantities a tilde:
$\tilde{e}_n(Q^2)$, $\tilde{Z}_{n,j}(Q^2,\mu^2)$,
and $\tilde{\Psi}_{n,j}(Q^2)$.
First of all, we realize that we have a different ``diagonal''
evolution
\begin{eqnarray}
 \label{eq:Power_Evo}
   \tilde{E}_n(Q^2,\mu^2)
   &=& \frac{\tilde{e}_n(Q^2)}{\tilde{e}_n(\mu^2)}\,;
   \qquad
   \tilde{e}_n(Q^2)
   \ =\ \Big[\frac{Q_0^2}{Q^2}\Big]^{\eta(n)},
  \\
  \label{eq:eta}
  \eta(n)
   &\equiv&
    \frac12\left(\frac{\alpha_s}{4\pi}\right)
     \left[\gamma_0(n)
         + \left(\frac{\alpha_s}{4\pi}\right)\gamma_1(n)
     \right] \, ,
\end{eqnarray}
where $Q_0^2$ is an arbitrary auxiliary scale.
Equations (\ref{eq:sol_Zn,n-k}), (\ref{eq:RG_a_n}), and
(\ref{eq:RG_Z_n,k}) remain valid,
but the ERBL\ equation for $a_n(Q^2)$ gets modified to assume the
following form
\begin{eqnarray}
 \frac{da_n(Q^2)}{d\ln(Q^2)}
   = - \eta(n) a_n(Q^2)
     + \frac{1}{2}
        \left(\frac{\alpha_s}{4\pi}\right)^2
         {\sum_{0\leq j<n}}'M_{j,n}a_{j}(Q^2)\,.
\end{eqnarray}
Equation (\ref{eq:ERBL-2loop:Z_nj}) then transforms to
\begin{eqnarray}
 \label{eq:ERBL-2loop_fix:Z_nj}
  \frac{d\tilde{Z}_{n,j}^\text{RG}(Q^2,\mu^2)}{d\ln Q^2}
   = \frac{1}{2}\left(\frac{\alpha_s}{4\pi}\right)^2
      \left[M_{j,n}\,\frac{\tilde{e}_j(Q^2)}{\tilde{e}_n(Q^2)}
         + {\sum_{j<k<n}}'M_{k,n}\,
             \frac{\tilde{e}_k(Q^2)}{\tilde{e}_n(Q^2)}\,
              \tilde{Z}_{k,j}^\text{RG}(Q^2,\mu^2)
      \right]
\end{eqnarray}
and for $j=n-2$ it reduces to
\begin{eqnarray}
 \label{eq:dZ/da_nn-2_fix}
  \frac{d\tilde{Z}_{n,n-2}^\text{RG}(Q^2,\mu^2)}{d\ln Q^2}
   = \frac{1}{2}\left(\frac{\alpha_s}{4\pi}\right)^2
      M_{n-2,n}\,\frac{\tilde{e}_{n-2}(Q^2)}{\tilde{e}_n(Q^2)}\,.
\end{eqnarray}
Its exact solution reads
\begin{eqnarray}
 \label{eq:Exact_Z_n,n-2_fix}
  \tilde{Z}_{n,n-2}^\text{RG}(Q^2,\mu^2)
   = \tilde{\Psi}_{n,n-2}^\text{RG}(Q^2)
   - \tilde{\Psi}_{n,n-2}^\text{RG}(\mu^2)
\end{eqnarray}
with
\begin{subequations}
\begin{eqnarray}
 \label{eq:RG_Psi_nn-2_fix}
  \tilde{\Psi}_{n,n-2}^\text{RG}(Q^2)
  &\equiv&
    \frac{1}{2}\left(\frac{\alpha_s}{4\pi}\right)^2\,
     \frac{L_{n,n-2}}{\eta(n)-\eta(n-2)}\,
      \frac{\tilde{e}_{n-2}(Q^2)}{\tilde{e}_n(Q^2)}\\
  &=&
    \left(\frac{\alpha_s}{4\pi}\right)\,
     \frac{L_{n,n-2}\ \left[Q^2/Q_0^2\right]^{\eta(n)-\eta(n-2)}}
          {\gamma_0(n)-\gamma_0(n-2)
          +\left[\alpha_s/(4\pi)\right]
           \left(\gamma_1(n)-\gamma_1(n-2)\right)}
       \,;~~~\label{eq:RG_Psi_nn-2_alpha}\\
 L_{n,n-2}
  &=& M_{n-2,n}\,.~~~\label{eq:RG_L_nn-2}
\end{eqnarray}
\end{subequations}
But for this solution we have already obtained a RG-improved
expression, given by (\ref{eq:RG_Z_n,k}).
After substituting this expression into the evolution equation
(\ref{eq:ERBL-2loop_fix:Z_nj}), we obtain
\begin{eqnarray}
 \label{eq:ERBL-2loop:Psi_nj_fix}
  \frac{d\tilde{\Psi}_{n,j}(Q^2)}{d\ln Q^2}
   = \frac{1}{2}\left(\frac{\alpha_s}{4\pi}\right)^2
      \left[M_{j,n}\,\frac{\tilde{e}_{j}(Q^2)}{\tilde{e}_n(Q^2)}
         + {\sum_{j<k<n}}' M_{k,n}
            \tilde{\Psi}_{k,j}(Q^2)
             \frac{\tilde{e}_{k}(Q^2)}
                  {\tilde{e}_n(Q^2)}
      \right]\,.
\end{eqnarray}
The solution of this differential equation can be represented as
\begin{subequations}
\begin{eqnarray}
 \label{eq:2loop_Psi_nj_fix}
  \tilde{\Psi}_{n,j}^\text{RG}(Q^2)
   &\equiv&
    \frac{1}{2}\left(\frac{\alpha_s}{4\pi}\right)^2\,
     \frac{L_{n,j}}{\eta(n)-\eta(j)}\
      \frac{\tilde{e}_{j}(Q^2)}
           {\tilde{e}_{n}(Q^2)}\\
  &=&
    \left(\frac{\alpha_s}{4\pi}\right)\,
     \frac{L_{n,j}\ \left[Q^2/Q_0^2\right]^{\eta(n)-\eta(j)}}
          {\gamma_0(n)-\gamma_0(j)
          +\left[\alpha_s/(4\pi)\right]
           \left(\gamma_1(n)-\gamma_1(j)\right)}
       \,,~~~\label{eq:RG_Psi_nj_alpha}
\end{eqnarray}
where the matrix $L_{n,j}$ is defined iteratively by
\begin{eqnarray}
 \label{eq:L_nj}
  L_{n,j}
   = M_{j,n}
   + \frac{1}{2}\left(\frac{\alpha_s}{4\pi}\right)^2\,
      {\sum_{j<k<n}}' M_{k,n}
       \frac{L_{k,j}}{\eta(k)-\eta(j)}\,.
\end{eqnarray}
\end{subequations}
In Fig.\ \ref{fig:comp_LM} we show numerical results for $L_{n,j}$
and $M_{j,n}$.
As one sees, the RG-improved solution $L_{n,j}$ is for $1\ll j\leq n-2$
3 to 4 times smaller compared to $M_{j,n}$.

Having obtained these exact solutions, it is instructive to see how the
result derived in~\cite{Mul95} for the asymptotic ($Q^2\to\infty$)
solution of the NLO ERBL\ equation with fixed $\alpha_s$ can be
reproduced.
From Eq.\ (\ref{eq:RG_a_n}) we know that the $Q^2$-dependence of the
Gegenbauer expansion coefficients $a_n$ is determined by the
combinations
$\tilde{e}_n(Q^2)\tilde{e}_{n}^{-1}(\mu^2)$
and
$\tilde{e}_n(Q^2)\tilde{Z}_{n,j}^\text{RG}(Q^2,\mu^2)
 \tilde{e}_{j}^{-1}(\mu^2)$.
The first term has an evident asymptotics:
\begin{eqnarray}
 \frac{\tilde{e}_n(Q^2)}{\tilde{e}_{n}^{-1}(\mu^2)}
 &=&
 \Big[\frac{\mu^2}{Q^2}\Big]^{\eta(n)}
 \ \stackrel{Q^2\to\infty}\rightarrow \
  \Big\{\begin{array}{cl}
    1, & \text{if~}n=0 \\
    0, & \text{if~}n\neq0
  \end{array}\,,
\end{eqnarray}
whereas the asymptotics of the second term can be determined as
follows.
First, let us determine the asymptotics of this term in the case
$j=j_2\equiv n-2$:
\begin{eqnarray}
 \tilde{e}_n(Q^2)\tilde{Z}_{n,j_2}^\text{RG}(Q^2,\mu^2)
 \tilde{e}_{j_2}(\mu^2)^{-1}
  = \frac{1}{2}\left(\frac{\alpha_s}{4\pi}\right)^2\,
     \left[\left(\frac{\mu^2}{Q^2}\right)^{\eta(j_2)}
          -\left(\frac{\mu^2}{Q^2}\right)^{\eta(n)}
     \right]\,
      \frac{L_{n,j_2}}{\eta(n)-\eta(j_2)}\,.~~~~\nonumber
\end{eqnarray}
For $Q^2\to\infty$ we have
\begin{eqnarray}
 \tilde{e}_n(Q^2)\tilde{Z}_{n,j_2}^\text{RG}(Q^2,\mu^2)
 \tilde{e}_{j_2}(\mu^2)^{-1}
 &\rightarrow&
  \left\{\begin{array}{cl}
    \Ds\frac{1}{2}\left(\frac{\alpha_s}{4\pi}\right)^2\,
     \frac{L_{n,0}}{\eta(n)}, & \text{if~}j_2=0 \\
    0, & \text{if~}j_2\neq0
        \end{array}\right.\,.
 \label{eq:asy_n-2}
\end{eqnarray}
The case $j=j_4\equiv n-4$ differs from the case $j=j_2$ by the
contribution of an additional term; viz.,
\begin{eqnarray}
 \left[\tilde{e}_n(Q^2)\,
       \tilde{Z}_{n,j_2}(Q^2,\mu^2)\,
       \tilde{e}_{j_2}(\mu^2)^{-1}
 \right]\,
  \tilde{\Psi}_{j_2,j_4}(\mu^2)\,
   \frac{\tilde{e}_{j_2}(\mu^2)}{\tilde{e}_{j_4}(\mu^2)}\,,
\end{eqnarray}
which vanishes as $Q^2\to\infty$ due to Eq.\ (\ref{eq:asy_n-2}) and
$j_2>j_4\geq0$.
The same conclusion can be drawn for all $j=n-k$ with $k\geq4$, so
that we can state that for $Q^2\to\infty$
\begin{eqnarray}
 \tilde{e}_n(Q^2)\tilde{Z}_{n,j}^\text{RG}(Q^2,\mu^2)
 \tilde{e}_{j}(\mu^2)^{-1}
 &\rightarrow&
  \left\{\begin{array}{cl}
         \Ds\left(\frac{\alpha_s}{4\pi}\right)\,
             \frac{L_{n,0}}{\gamma_0(n)+(\alpha_s/4\pi)\gamma_1(n)},
            & \text{if~}j=0 \\
         0, & \text{if~}j\neq0
         \end{array}
  \right.\,.
 \label{eq:asy_n,j}
\end{eqnarray}
Then, in accordance with Eq.\ (\ref{eq:RG_a_n}), the new
(non-polynomial) asymptotic distribution amplitude is given by
\begin{eqnarray}
 \label{eq:NLO_AsDA}
  \varphi^\text{NLO}_\text{as}(x;\alpha_s)
   &=&
    \varphi_\text{as}(x)
     \left[1
         + \left(\frac{\alpha_s}{4\pi}\right)\,
            {\sum_{n\geq2}}'
             \frac{L_{n,0}}{\gamma_0(n)+(\alpha_s/4\pi)\gamma_1(n)}\,
              C_{n}^{3/2}(2x-1)
     \right]\,.
\end{eqnarray}
Expanding this expression in $\alpha_s$ and retaining only the
$O(\alpha_s)$-terms, we obtain
\begin{eqnarray}
 \label{eq:NLO_AsDA_LO+NLO}
  \varphi^\text{NLO}_\text{as}(x;\alpha_s)
   &=&
    \varphi_\text{as}(x)
     \left[1
         + \left(\frac{\alpha_s}{4\pi}\right)\,
            {\sum_{n\geq2}}'
             \frac{M_{0,n}}{\gamma_0(n)}\,
              C_{n}^{3/2}(2x-1)
     \right]
     + O(\alpha_s^2)\,.
\end{eqnarray}
\begin{figure}[h]
\centerline{\includegraphics[width=\textwidth]{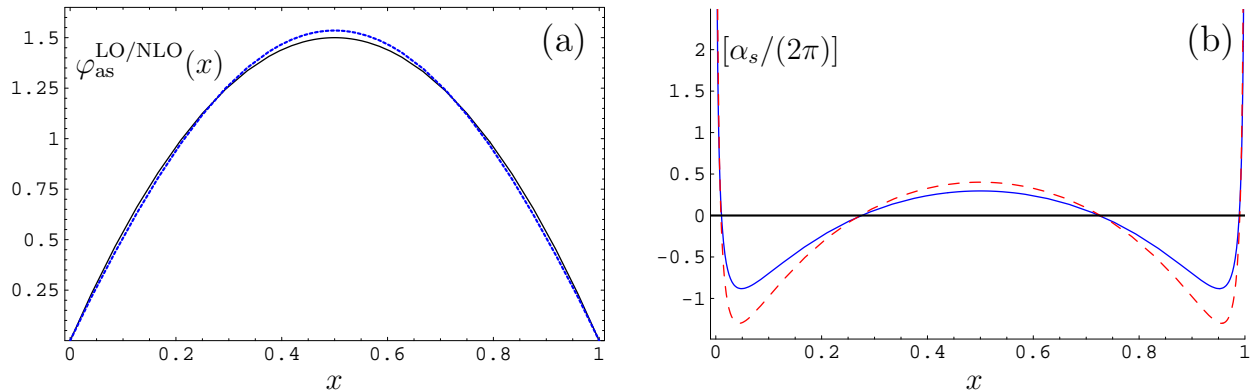}}%
\caption{\label{fig:comp_DA_Fix}\footnotesize
    Evolution of the pion DA for fixed $\alpha_s=0.5$
    and three active flavors.
    As a nonperturbative input at $\mu_0^2=0.25~\text{GeV}^2$
    we take the asymptotic DA $\varphi_\text{as}(x)$
    in the leading order.
    (a) The LO asymptotic DA, $\varphi_\text{as}(x)$,
    (solid line) in comparison with the exact NLO asymptotic DA
    $\varphi_\text{as}^\text{NLO}(x,\alpha_s)$ (Eq.\
    (\ref{eq:NLO_AsDA}), dashed line).
    (b) The relative NLO corrections
    to $\varphi_\text{as}(x,Q^2)$ at $Q^2=4~\text{GeV}^2$
    are shown in units of $\alpha_s/(2\pi)$ following~\cite{Mul95} for
    an easier comparison: the exact RG prediction,
    Eqs.\ (\ref{eq:2loop_Psi_nj_fix})--(\ref{eq:L_nj}),
    is denoted by a solid line and the result of \cite{Mul95},
    Eq.\ (\ref{eq:NLO_AsDA_LO+NLO_res}), is represented by a dashed
    line.
    }
\end{figure}

Using the explicit form of $M_{0,n}$---see Eqs.\
(\ref{eq:Exact_Mkn})--(\ref{eq:Akn})---this expression can be recast in
the form
\begin{subequations}
\begin{eqnarray}
 \label{eq:NLO_AsDA_LO+NLO_res}
  \varphi^\text{NLO}_\text{as}(x;\alpha_s)
   &=&
    \varphi_\text{as}(x)
     \left[1
         + \frac{\alpha_s}{4\pi}\,
            \left(C_F \phi_F(x) + b_0 \phi_{b_0}(x)\right)
     \right]
     + O(\alpha_s^2)
\end{eqnarray}
which coincides with Eq.\ (31) derived in Ref.\ \cite{Mul95}.
Here, the following abbreviations have been introduced
\begin{eqnarray}
 \phi_F(x)
   &\equiv&
    {\sum_{n\geq2}}'
     \frac{4\,(2n+3)}{(n+1)(n+2)}\,
      \left[\frac{4A_{n0}}{n(n+3)}
           + A_{n0} - \psi(n+2) + \psi(1)
      \right]\,
      C_{n}^{3/2}(2x-1)~~~~\nonumber\\
   &=& \ln^2\left[\frac{1}{x}-1\right]
     + 2 - \frac{\pi^2}{3}\,;\\
 \phi_{b_0}(x)
   &\equiv&
    {\sum_{n\geq2}}'
     \frac{-4\,(2n+3)}{n(n+1)(n+2)(n+3)}\,
      C_{n}^{3/2}(2x-1)
  \ =\ \ln\left[x(1-x)\right]
      + \frac{5}{3}\,.
  \label{eq:phi_b0}
\end{eqnarray}
\end{subequations}

Hence, we conclude that in the fixed $\alpha_s$ case our approach
fully reproduces the two-loop result of Ref.\ \cite{Mul95} and the
part (\ref{eq:phi_b0}) coincides with Mikhailov's finding in
\cite{MS00}.
But our approach contains more information.
Because of the RG improvement, Eq.\ (\ref{eq:NLO_AsDA}) contains the
coefficients $L_{n,0}$ instead of $M_{0,n}$ entering this equation in
the approach of~\cite{Mul95}---cf.\ Eq.\ (\ref{eq:NLO_AsDA_LO+NLO}).
The advantage of the coefficients $L_{n,0}$ is that they are much
smaller, as we clearly see from Fig.\ \ref{fig:comp_LM} by comparing
the two approaches.
In Fig.\ \ref{fig:comp_DA_Fix} we show the comparison of the two
approaches for the pion DA itself.
Note that in our analysis we use the same numerical values
$\alpha_s=0.5$, $N_f=3$,
$\mu^2=0.25~\text{GeV}^2$,
and
$Q^2=4~\text{GeV}^2$ as in~\cite{Mul95} and take into account in
Eq.\ (\ref{eq:RG_a_n}) the first 100 nontrivial terms
$(n=2,4,\ldots,200)$ to approximate the evolved distribution amplitude.
It is important to stress that our results, obtained in the
\MS scheme of perturbative QCD for a fixed coupling, can be
related to those derived before by M\"{u}ller \cite{Mue98,Mue99}
in the conformally covariant subtraction (CS) scheme.
More specifically, the solutions of the ERBL\ evolution equation in
each scheme can be interlinked in the conformal limit ($\beta=0$) on
account of a finite refactorization \cite{Mue98}.
This means that our solution (\ref{eq:NLO_AsDA_LO+NLO_res}) in the
\MS scheme and M\"{u}ller's corresponding results in the
CS\ scheme \cite{Mue98,Mue99} are connected by a RG transformation
matrix determined by the special conformal anomaly \cite{BDFL86}.

\section{Conclusions}
In this paper we analyzed the ERBL\ evolution equation for the meson
distribution amplitude at the two-loop level.
Our main conclusions can be summarized as follows.
\begin{itemize}
\item We worked out a procedure to solve the ERBL\ evolution equation
in the NLO approximation, Eqs.\ (\ref{eq:RG_a_n}), (\ref{eq:RG_Z_n,k}),
which ensures the RG properties.

\item Using this method, we obtained a RG-improved solution of the NLO
ERBL\ evolution equation in the form of quadratures,
Eqs.\ (\ref{eq:2loop_Psi_nj}), (\ref{eq:RGimp_Phi_nk}).

\item We worked out an approximate version of our procedure, given by
Eqs.\ (\ref{eq:Appro_Psi_nk}) and (\ref{eq:Appro_Z_nk}), which is
completely analytical.
Its accuracy is rather high with an uncertainty of the order of only a
few \%, except for the case of $Z_{4,0}$, for which the error is about
$-10$\%.

\item We analyzed the importance of the NLO evolution for the inverse
moment of the meson distribution amplitude.
We confirmed the conclusion drawn in~\cite{BPSS04} that it is possible
to use the LO evolution for this quantity with the induced overall
error being smaller than 1\%.

\end{itemize}
The most discernible theoretical result of this investigation is that
the RG-improved evolution is slower and the higher harmonics stronger
suppressed relative to the approximate MR scheme, enhancing the
self-consistency of QCD perturbation theory.

\begin{acknowledgments}
This work was supported in part by the Russian Foundation for
Fundamental Research
(grants 03-02-16816, 03-02-04022 and 05-01-00992),
the Heisenberg--Landau Programme,
and the Deutsche Forschungsgemeinschaft (DFG)
(grant 436 RUS 113/752/0-1).
We are particularly grateful to S.~V.~Mikhailov for stimulating
discussions that inspired us to investigate this subject.
We would also like to thank to K.~Passek-Kumeri\v{c}ki
for useful discussions and remarks.
One of us (A.P.B.) is grateful to Prof.\ Klaus Goeke for the warm
hospitality at Bochum University, where part of this work was carried
out.
\end{acknowledgments}

\newpage
\begin{appendix}
\appendix
\section{Beta-function and anomalous dimensions}
 \renewcommand{\theequation}{\thesection.\arabic{equation}}
  \label{sec:App-A}\setcounter{equation}{0}
The standard $\beta$-function coefficients are given by
\begin{eqnarray}
 \label{eq:beta-b01,c1}
  b_0 = \frac{11 N_c - 2 N_f}{3}\,;\quad
  b_1 = \frac{34}{3}N_c^2
        - \left(2C_\text{F} + \frac{10}{3}N_c\right)N_f\,.
\end{eqnarray}
To one-loop, the anomalous dimensions read
\begin{eqnarray}
  \gamma_0(n)
  = 2C_F\left[4\sum_{i=1}^{n+1}\frac{1}{i}
            - 3 - \frac{2}{(n+1)(n+2)}
        \right]\,,
\end{eqnarray}
whereas the anomalous dimensions for arbitrary $n$ at the two-loop
order may be found in~\cite{Flo77,GonArr79}, with
\begin{eqnarray}
 \gamma_1(0)\ =\ 0\,, \quad \quad
 \gamma_1(2) &=& \frac{34450}{243}-\frac{830}{81}N_f\,, \qquad
 \gamma_1(4)\ =\ \frac{662846}{3375}-\frac{31132}{2025}N_f\,.
 \label{eq:gamma_04}
\end{eqnarray}
The normalization of  the Gegenbauer polynomials $\psi_n(x)$
with respect to the weight function $\Omega(x)=6\,x\,(1-x)$ is
\begin{eqnarray}
  N_n &=& \int_0^1\Omega(x)\psi_n^2(x) dx
     \ =\ \frac{3(n+1) (n+2)}{2(2n+3)}\,.
  \label{eq:N_n}
\end{eqnarray}
The values of the first few matrix elements $M_{j,n}$ have been
calculated numerically in~\cite{MR86ev} to be
\begin{eqnarray}
 M_{0 2} = -11.2 + 1.73  N_f,~~
 M_{0 4} = -1.41 + 0.565 N_f, ~~
 M_{2 4} = -22.0 + 1.65  N_f\,.
\end{eqnarray}
In particular, for $N_f=3$, they read
\begin{eqnarray}
 \label{eq:M024}
 M_{0 2} = -6.01,~~
 M_{0 4} = 0.285, ~~
 M_{2 4} = -17.05\,.
\end{eqnarray}
It is worth pointing out here
that M\"{u}ller in Ref.\ \cite{Mul94}
has obtained analytic expressions for the matrix elements $M_{k,n}$
for all values $j=0, 2, \ldots < n=2, 4, \ldots$
\footnote{%
The function $\psi(z)$ is defined as usual by
$\psi(z) = \frac{d}{dz}\ln\Gamma(z)$.}
\begin{eqnarray}
  M_{j,n}
   &=& 2\,\frac{N_j}{N_n}\,
        C_{n,j}^{(1)}\,
         \big[\gamma_{0}(n)-\gamma_{0}(j)\big]\,,
 \label{eq:Exact_Mkn}\\
   C_{n,j}^{(1)}
    &=& (2 j + 3)
           \left[\frac{-\gamma_{0}(j)-2 b_0+8 C_F A_{n,j}}
                      {2 (n-j)(n+j+3)}
               + \frac{2 C_F (A_{n,j} - \psi (n+2) + \psi(1) )}
                      {(j+1)(j+2)}
           \right]\,, \label{eq:Ckn1} \\
   A_{n,j}
   &=& \psi\left(\frac{n+j+4}{2}\right)
     - \psi\left(\frac{n-j}{2}\right)
     + 2 \psi(n-j) - \psi(n+2)- \psi(1)\,.
 \label{eq:Akn}
\end{eqnarray}

\section{Proof of E\lowercase{q}.\ (\ref{eq:sol_Zn,n-k})}
 \renewcommand{\theequation}{\thesection.\arabic{equation}}
  \label{sec:App-MI}\setcounter{equation}{0}
Using the principle of mathematical induction we want to prove that
\begin{eqnarray}
 Z_{n,n-k}(Q^2,\mu^2) = \Psi_{n,n-k}(Q^2)-\Psi_{n,n-k}(\mu^2)
 - {\sum_{0<j<k}}'Z_{n,n-j}(Q^2,\mu^2)\Psi_{n-j,n-k}(\mu^2)
\end{eqnarray}
is the solution of Eq.\ (\ref{eq:RG_Zn,n-k}).

\textbf{\fbox{1}} We proved this for $k=2$.

\textbf{\fbox{2}} Let us assume that it is valid for all $k \leq m-2$
and then prove that it is valid for $k = m$.
We have\\
\begin{eqnarray}
 Z(m,Q^2,q^2,\mu^2)
  &\equiv& - Z_{n,n-m}(Q^2,\mu^2) + Z_{n,n-m}(Q^2,q^2)
           + Z_{n,n-m}(q^2,\mu^2)\ =\
\qquad\qquad\qquad\nonumber\\
  &=& {\sum_{0<j<m}}'Z_{n,n-j}(Q^2,\mu^2)\Psi_{n-j,n-m}(\mu^2)
\nonumber\\
  &-& {\sum_{0<j<m}}'
       \left[Z_{n,n-j}(Q^2,q^2)\Psi_{n-j,n-m}(q^2)
           + Z_{n,n-j}(q^2,\mu^2)\Psi_{n-j,n-m}(\mu^2)
       \right]\ = \nonumber\\
  &=& {\sum_{0<j<m}}'
      \left[Z_{n,n-j}(Q^2,\mu^2)
          - Z_{n,n-j}(Q^2,q^2)
          - Z_{n,n-j}(q^2,\mu^2)\right]\Psi_{n-j,n-m}(\mu^2)
\nonumber\\
  &+& {\sum_{0<j<m}}'Z_{n,n-j}(Q^2,q^2)
      \left[\Psi_{n-j,n-m}(q^2) - \Psi_{n-j,n-m}(\mu^2)\right]
\nonumber\,.
\end{eqnarray}
Now we can use Eq.\ (\ref{eq:sol_Zn,n-k}) to rewrite the last line in
the previous equation in the following way (on account of $j\leq m-2$
in all sums)
\begin{eqnarray}
 \Psi_{n-j,n-m}(q^2) - \Psi_{n-j,n-m}(\mu^2)
  = Z_{n-j,n-m}(q^2,\mu^2)
  + {\sum_{j<k<m}}'Z_{n-j,n-k}(q^2,\mu^2)\Psi_{n-k,n-m}(\mu^2)
\nonumber\,.
\end{eqnarray}
Then we find
\begin{eqnarray}
 Z(m,Q^2,q^2,\mu^2)
  &=& {\sum_{0<j<m}}'
      \left[Z_{n,n-j}(Q^2,\mu^2)
          - Z_{n,n-j}(Q^2,q^2)
          - Z_{n,n-j}(q^2,\mu^2)\right]\Psi_{n-j,n-m}(\mu^2)
\nonumber\\
  &+& {\sum_{0<j<m}}'Z_{n,n-j}(Q^2,q^2)Z_{n-j,n-m}(q^2,\mu^2)
\nonumber\\
  &+& {\sum_{0<k<m}}'{\sum_{k<j<m}}'
       Z_{n,n-k}(Q^2,q^2)Z_{n-k,n-j}(q^2,\mu^2)\Psi_{n-j,n-m}(\mu^2)
\nonumber\,,
\end{eqnarray}
where in the last line  we have introduced new indices ($j,k \to k,j$).
Changing the order of summations in the double sum gives
\begin{eqnarray}
 Z(m,Q^2,q^2,\mu^2)
  &=& {\sum_{0<j<m}}'
      \left[Z_{n,n-j}(Q^2,\mu^2)
          - Z_{n,n-j}(Q^2,q^2)
          - Z_{n,n-j}(q^2,\mu^2)\right]\Psi_{n-j,n-m}(\mu^2)
\nonumber\\
  &+& {\sum_{0<j<m}}'Z_{n,n-j}(Q^2,q^2)Z_{n-j,n-m}(q^2,\mu^2)
\nonumber\\
  &+& {\sum_{0<j<m}}'{\sum_{0<k<j}}'
       Z_{n,n-k}(Q^2,q^2)Z_{n-k,n-j}(q^2,\mu^2)\Psi_{n-j,n-m}(\mu^2)\ =
\nonumber
\end{eqnarray}
\begin{eqnarray}
  &=& {\sum_{0<j<m}}'
      \Big[Z_{n,n-j}(Q^2,\mu^2)
          - Z_{n,n-j}(Q^2,q^2)
          - Z_{n,n-j}(q^2,\mu^2)\nonumber\\
  & & \qquad \ \
          + {\sum_{0<k<j}}'Z_{n,n-k}(Q^2,q^2)Z_{n-k,n-j}(q^2,\mu^2)
      \Big]\Psi_{n-j,n-m}(\mu^2)\nonumber\\
  &+& {\sum_{0<j<m}}'Z_{n,n-j}(Q^2,q^2)Z_{n-j,n-m}(q^2,\mu^2)
\nonumber\,.
\end{eqnarray}
By virtue of $j\leq m-2$ in all sums, we may use Eq.\
(\ref{eq:RG_Zn,n-k}) for the expression inside the square brackets to
show that it is identically equal to zero.
Therefore, we find
\begin{eqnarray}
 Z(m,Q^2,q^2,\mu^2)
  &\equiv& Z_{n,n-m}(Q^2,\mu^2)-Z_{n,n-m}(Q^2,q^2)
         - Z_{n,n-m}(q^2,\mu^2)
\nonumber\\
  &=& {\sum_{0<j<m}}'Z_{n,n-j}(Q^2,q^2)Z_{n-j,n-m}(q^2,\mu^2)
\nonumber\,.
\end{eqnarray}
This way we obtain just the desired expression, cf.\
Eq.\ (\ref{eq:RG_Zn,n-k}).
\hfill{\textbf{\fbox{Q.E.D.}}}

\end{appendix}


\newcommand{\noopsort}[1]{} \newcommand{\printfirst}[2]{#1}
  \newcommand{\singleletter}[1]{#1} \newcommand{\switchargs}[2]{#2#1}

\end{document}